\documentclass[a4paper,11pt]{article}

\usepackage{jheppub_mod} 
\usepackage[T1]{fontenc} 

\usepackage{amsmath,booktabs}
\usepackage{graphicx}
\usepackage{xspace}
\usepackage{color}
\pagecolor{white}
\usepackage[dvipsnames]{xcolor}
\usepackage{url}
\usepackage[utf8]{inputenc}
\usepackage{epstopdf}
\usepackage{hyperref}
\usepackage{multirow}
\usepackage{gensymb}

\usepackage{ifmtarg}

\newcommand{\mpi}{M_\pi}

\newcommand{\beq}{\begin{equation}}
\newcommand{\eeq}{\end{equation}}

\newcommand{\Order}{\mathcal{O}}

\newcommand{\GeV}{\,\text{GeV}}
\newcommand{\MeV}{\,\text{MeV}}

\renewcommand{\Im}{\text{Im}\,}

\newcommand{\F}{\mathcal{F}}

\newcommand{\mA}{M_A}

\newcommand{\sm}{s_\text{m}}
\newcommand{\Mr}{M_\rho}
\newcommand{\Meff}{M_\text{eff}}
\newcommand{\Ma}{M_{a_1}}
\newcommand{\Mf}{M_{f_1}}
\newcommand{\Mfp}{M_{f_1'}}

\makeatletter
\newcommand{\Cr}[2]{\@ifmtarg{#2}{\mathcal{C}_{#1}}{\mathcal{C}_{#1}\big[#2\big]}}
\makeatother

\allowdisplaybreaks[1]

\title{Dispersion relation for hadronic light-by-light scattering: subleading contributions}

\author[a]{Martin Hoferichter,}
\author[b,c]{Peter Stoffer,}
\author[a]{and Maximilian Zillinger}

\affiliation[a]{Albert Einstein Center for Fundamental Physics, Institute for Theoretical Physics, University of Bern, Sidlerstrasse 5, 3012 Bern, Switzerland}

\affiliation[b]{Physik-Institut, Universit\"at Z\"urich, Winterthurerstrasse 190, 8057 Z\"urich, Switzerland}

\affiliation[c]{PSI Center for Neutron and Muon Sciences, 5232 Villigen PSI, Switzerland}

\preprint{PSI-PR-24-26, ZU-TH 60/24}

\emailAdd{hoferichter@itp.unibe.ch}
\emailAdd{stoffer@physik.uzh.ch}
\emailAdd{zillinger@itp.unibe.ch}

\abstract{In this work, we present an evaluation of subleading effects  in the hadronic light-by-light contribution to the anomalous magnetic moment of the muon. Using a recently derived optimized basis, we first study the matching of axial-vector contributions to short-distance constraints at the level of the scalar basis functions, finding that also the tails of the pseudoscalar poles and tensor mesons play a role. We then develop a matching strategy that allows for a combined evaluation of axial-vector and short-distance constraints, supplemented by an estimate of tensor-meson contributions based on simplified assumptions for their transition form factors. Uncertainties are primarily propagated from the axial-vector transition form factors and the variation of the matching scale, but we also consider estimates of the low-energy effect of hadronic states not explicitly included. In total, we obtain $a_\mu^\text{HLbL}\big|_\text{subleading}=33.2(7.2)\times 10^{-11}$, which in combination with previously evaluated contributions in the dispersive approach leads to $a_\mu^\text{HLbL}\big|_\text{total}=101.9(7.9)\times 10^{-11}$.}

\begin{document}
\maketitle
	
\section{Introduction}
\label{sec:intro}

Hadronic light-by-light (HLbL) scattering constitutes a key contribution to the Standard-Model prediction of the muon anomalous magnetic moment $a_\mu$~\cite{Aoyama:2020ynm,Aoyama:2012wk,Aoyama:2019ryr,Czarnecki:2002nt,Gnendiger:2013pva,Davier:2017zfy,Keshavarzi:2018mgv,Colangelo:2018mtw,Hoferichter:2019gzf,Davier:2019can,Keshavarzi:2019abf,Hoid:2020xjs,Kurz:2014wya,Melnikov:2003xd,Colangelo:2014dfa,Colangelo:2014pva,Colangelo:2015ama,Masjuan:2017tvw,Colangelo:2017qdm,Colangelo:2017fiz,Hoferichter:2018dmo,Hoferichter:2018kwz,Gerardin:2019vio,Bijnens:2019ghy,Colangelo:2019lpu,Colangelo:2019uex,Blum:2019ugy,Colangelo:2014qya}. While, at present, the main limitation in the prediction derives from tensions in hadronic vacuum polarization~\cite{Colangelo:2022jxc}, both within data-driven approaches~\cite{Davier:2017zfy,Keshavarzi:2018mgv,Colangelo:2018mtw,Hoferichter:2019gzf,Davier:2019can,Keshavarzi:2019abf,Hoid:2020xjs,Crivellin:2020zul,Keshavarzi:2020bfy,Malaescu:2020zuc,Colangelo:2020lcg,Stamen:2022uqh,Colangelo:2022vok,Colangelo:2022prz,Hoferichter:2023sli,Hoferichter:2023bjm,Stoffer:2023gba,Davier:2023fpl,CMD-3:2023alj,CMD-3:2023rfe,Leplumey:2025kvv} (emphasizing the importance of radiative corrections~\cite{Campanario:2019mjh,Ignatov:2022iou,Colangelo:2022lzg,Monnard:2021pvm,Abbiendi:2022liz,BaBar:2023xiy,Aliberti:2024fpq}) and with lattice QCD~\cite{Borsanyi:2020mff,Ce:2022kxy,ExtendedTwistedMass:2022jpw,FermilabLatticeHPQCD:2023jof,RBC:2023pvn,Boccaletti:2024guq,Blum:2024drk,Djukanovic:2024cmq,Bazavov:2024eou}, also the HLbL contribution deserves further attention when compared to the current experimental world average dominated by the Fermilab experiment~\cite{Muong-2:2023cdq,Muong-2:2024hpx}
\beq
a_\mu^\text{exp}=116\,592\,059(22)\times 10^{-11},
\eeq
especially in view of the upcoming final result, which is likely to even exceed the original precision goal~\cite{Muong-2:2015xgu}. At this level, improvement by at least another factor two is required for
  $a_\mu^\text{HLbL} = 92(19) \times 10^{-11}$~\cite{Aoyama:2020ynm,Melnikov:2003xd,Masjuan:2017tvw,Colangelo:2017qdm,Colangelo:2017fiz,Hoferichter:2018dmo,Hoferichter:2018kwz,Gerardin:2019vio,Bijnens:2019ghy,Colangelo:2019lpu,Colangelo:2019uex,Pauk:2014rta,Danilkin:2016hnh,Jegerlehner:2017gek,Knecht:2018sci,Eichmann:2019bqf,Roig:2019reh} to fully profit from the experimental gain in precision. Recent developments concern improved evaluations in lattice QCD~\cite{Blum:2019ugy,Chao:2021tvp,Chao:2022xzg,Blum:2023vlm,Fodor:2024jyn} and various
  phenomenological studies aimed at different aspects of HLbL scattering.

  In a dispersive approach~\cite{Colangelo:2014dfa,Colangelo:2014pva,Colangelo:2015ama,Hoferichter:2013ama,Colangelo:2017fiz}, one aims at reconstructing the HLbL tensor in terms of its singularities, which, in turn, can be related to experimental observables. In most cases, those observables again need to be reconstructed from dispersion relations to cover the entire range of photon virtualities, a program which has been completed for the $\pi^0$ pole~\cite{Schneider:2012ez,Hoferichter:2012pm,Hoferichter:2014vra,Hoferichter:2018dmo,Hoferichter:2018kwz,Hoferichter:2021lct}, $\eta$, $\eta'$ poles~\cite{Stollenwerk:2011zz,Hanhart:2013vba,Kubis:2015sga,Holz:2015tcg,Holz:2022hwz,Holz:2022smu,Holz:2024lom,Holz:2024diw}, pion- and kaon-box contributions~\cite{Colangelo:2017qdm,Colangelo:2017fiz,Stamen:2022uqh}, and unitarity corrections that can be interpreted as effects related to the light scalars $f_0(500)$~\cite{Colangelo:2017qdm,Colangelo:2017fiz}, $f_0(980)$~\cite{Danilkin:2021icn}, and $a_0(980)$~\cite{Deineka:2024mzt}. For heavier intermediate states, such as axial-vector contributions, detailed studies of the respective transition form factors (TFFs) have been performed~\cite{Hoferichter:2020lap,Zanke:2021wiq,Hoferichter:2023tgp,Ludtke:2024ase}, and a new HLbL basis derived that is optimized for the evaluation of axial-vector resonances, while leaving the already dispersively evaluated parts unaltered~\cite{Hoferichter:2024fsj}. Moreover, an alternative dispersive formalism in soft-photon (triangle) kinematics has been developed, both for the HLbL tensor~\cite{Ludtke:2023hvz} and the vector--vector--axial-vector ($VVA$) correlator~\cite{Ludtke:2024ase}, primarily to be able to evaluate tensor-meson contributions beyond the approximation found in Ref.~\cite{Hoferichter:2024fsj} and to better assess uncertainties related to the truncation of intermediate states. In this work, describing the details of Ref.~\cite{Hoferichter:2024vbu}, we aim at providing a complete dispersive evaluation of the HLbL contribution that reflects these developments to the furthest extent possible at this point.

  To this end, a key aspect concerns the matching to short-distance constraints (SDCs), subleading corrections to which have become available in recent years~\cite{Bijnens:2020xnl,Bijnens:2021jqo,Bijnens:2022itw,Bijnens:2024jgh}, allowing for a robust estimate of the corresponding uncertainties. In addition, the implementation of SDCs, especially regarding the role of axial-vector states, has received lots of attention~\cite{Knecht:2020xyr,Masjuan:2020jsf,Ludtke:2020moa,Colangelo:2021nkr}, including holographic models of QCD~\cite{Leutgeb:2019gbz,Cappiello:2019hwh,Leutgeb:2021mpu,Leutgeb:2022lqw,Colangelo:2024xfh,Leutgeb:2024rfs} and Dyson--Schwinger equations~\cite{Eichmann:2024glq}. This interplay with SDCs is of particular concern, given that the dominant uncertainty in Ref.~\cite{Aoyama:2020ynm} reflected ambiguities and model dependences related to the evaluation of subleading hadronic states themselves, but also, crucially, the transition to the asymptotic region.

  To address these points, we pursue the following strategy. Below a matching scale $Q_0$ we evaluate axial-vector contributions in the basis of Ref.~\cite{Hoferichter:2024fsj}, with TFFs from Refs.~\cite{Hoferichter:2020lap,Zanke:2021wiq,Hoferichter:2023tgp}, see Sec.~\ref{sec:narrow_resonance}. We also estimate corrections from heavy scalars and tensors, and include mass effects in the axial-vector TFFs, see App.~\ref{app:mass_effects}. Above $Q_0$, we use the perturbative result from Ref.~\cite{Bijnens:2021jqo}.
  In parts of the mixed regions, the operator product expansion (OPE) allows one to match to the $VVA$ correlator, for which we can rely on the dedicated calculation from Ref.~\cite{Ludtke:2024ase}, see App.~\ref{app:VVA_matching}. In Sec.~\ref{sec:matching_SDCs}, we study to which extent the sum of hadronic states indeed matches onto the SDCs, finding overall good agreement, but not a perfect matching. Apart from experimental inputs, our main estimate of the uncertainty therefore concerns the variation of the matching scale $Q_0$, but we also study the possible impact of missed heavier states by improving the matching using effective poles with couplings determined from the asymptotic limits, see App.~\ref{app:effective_poles}. The consequences for $a_\mu$ are detailed in Sec.~\ref{sec:amu}, including the dependence on the matching scale $Q_0$, before we conclude in Sec.~\ref{sec:summary}.

\section{Narrow-resonance contributions}
\label{sec:narrow_resonance}

To describe the contributions from hadronic states we rely on a narrow-resonance approximation, using the optimized basis for the HLbL tensor from Ref.~\cite{Hoferichter:2024fsj}. In this basis, all states with angular momentum $J\leq 1$ can be evaluated without the appearance of kinematic singularities. Moreover, also the tensor-meson scalar functions become free of kinematic singularities if only the TFF $\F_1^T$ is non-zero, as predicted by the quark model~\cite{Schuler:1997yw}, and we will use this simplified scenario as a way to obtain a first estimate of tensor-meson contributions within our dispersive framework. For explicit expressions for the relevant scalar functions we refer to Ref.~\cite{Hoferichter:2024fsj}, whose notation we will follow throughout this paper.

While light scalar resonances are covered by the $S$-wave rescattering contributions as evaluated dispersively in Refs.~\cite{Colangelo:2017qdm,Colangelo:2017fiz,Danilkin:2021icn,Deineka:2024mzt}, the potential impact of heavier states, such as $f_0(1370)$ and $a_0(1450)$, was studied in Ref.~\cite{Danilkin:2021icn} in a narrow-resonance approximation as well. We reproduce these numbers in
Table~\ref{tab:narrow_resonance_scalar_tensor}, imposing, in addition, a cut in the photon virtualities $Q_i\leq Q_0$ ($q_i^2=-Q_i^2$), where the variation of the matching scale $Q_0$ will play a central role in assessing the final uncertainties. For the TFFs we take the form~\cite{Schuler:1997yw}
\beq
\frac{\F_1^S(q_1^2,q_2^2)}{\F_1^S(0,0)}=\frac{\Lambda_S^2(3\Lambda_S^2-q_1^2-q_2^2)}{3(\Lambda_S^2-q_1^2-q_2^2)^2},\qquad
\frac{\F_2^S(q_1^2,q_2^2)}{\F_1^S(0,0)}=-\frac{2\Lambda_S^4}{3(\Lambda_S^2-q_1^2-q_2^2)^2},
\eeq
with normalization $\F_1^S(0,0)$ determined from the two-photon widths $\Gamma_{\gamma\gamma}$ as discussed in Ref.~\cite{Danilkin:2021icn}. We identify the scale as $\Lambda_S=\Mr$, motivated by the observation that for $f_0(980)$ and $a_0(980)$ the expected vector-meson-dominance (VMD) scale indeed leads to better agreement with the explicit calculation in terms of helicity partial waves~\cite{Danilkin:2021icn,Deineka:2024mzt}.

\begin{table}[t]
	\centering
	\renewcommand{\arraystretch}{1.3}
	\begin{tabular}{lrrrrr}
	\toprule
	Resonance & $M \ [\text{GeV}]$ & $\Gamma_{\gamma\gamma}\ [\text{keV}]$ & $a_\mu[\bar\Pi_{1,2}] \ [10^{-11}]$ & $a_\mu[\bar\Pi_{3\text{--}12}] \ [10^{-11}]$ & Sum $[10^{-11}]$\\\midrule
	$f_0(1370)$ & $1.35(15)$ & $3.0(1.5)$ & -- & $-0.6(3)$ & $-0.6(3)$\\
	$a_0(1450)$ & $1.439(34)$ & $1.0(5)$ & -- & $-0.2(1)$ & $-0.2(1)$\\
	$f_0(1500)$ & $1.522(25)$ & not seen & -- & -- & --\\\midrule
	$S$ sum & & & -- & $-0.7(3)$ & $-0.7(3)$\\\midrule
	$f_2(1270)$ & $1.2754(8)$ & $2.65(45)$ & $1.9(3)$ & $-3.8(7)$ & $-1.9(3)$\\
	$a_2(1320)$ & $1.3182(6)$ & $1.01(9)$ & $0.6(1)$ & $-1.2(1)$ & $-0.6(1)$\\
	$f_2'(1525)$ & $1.5173(24)$ & $0.08(2)$ & -- & -- & --\\\midrule
	$T$ sum & &  & $2.6(3)$ & $-5.1(7)$ & $-2.5(3)$\\
\bottomrule
	\renewcommand{\arraystretch}{1.0}
	\end{tabular}
	\caption{Scalar and tensor narrow-resonance contributions for $Q_0=1.5\GeV$. The scale in the TFFs it set to $M_\rho$, see main text, and the quoted errors represent the uncertainty in the two-photon coupling. For the scalars, the ranges for $\Gamma_{\gamma\gamma}$ are chosen following the discussion in Ref.~\cite{Danilkin:2021icn}, while for the tensors we take over the values from Ref.~\cite{ParticleDataGroup:2024cfk}. The contributions from the $f_2'(1525)$ are below $0.1\times 10^{-11}$ and therefore omitted.}
	\label{tab:narrow_resonance_scalar_tensor}
\end{table}

Similarly, for the tensor states $f_2(1270)$, $a_2(1320)$, and $f_2'(1525)$ we use
\beq
\label{TFF_tensor}
\frac{\F_1^T(q_1^2,q_2^2)}{\F_1^T(0,0)}=\bigg(\frac{\Lambda_T^2}{\Lambda_T^2-q_1^2-q_2^2}\bigg)^2,\qquad \F_{2\text{--}5}^T(q_1^2,q_2^2)=0,
\eeq
with $\Lambda_T=\Mr$. In this case, the VMD scale is further motivated by the fact that the $f_2(1270)$ arises primarily as a unitarization of a vector-meson left-hand cut in $\gamma^*\gamma^*\to\pi\pi$~\cite{Garcia-Martin:2010kyn,Hoferichter:2011wk,Moussallam:2013una,Hoferichter:2013ama,Danilkin:2018qfn,Hoferichter:2019nlq,Danilkin:2019opj} (see also Refs.~\cite{Lu:2020qeo,Schafer:2023qtl,Deineka:2024mzt} for the $\pi\eta$ case), and the relevant scale in the TFFs that determine the dependence on the photon virtualities is again set by $\Mr$. The results shown in Table~\ref{tab:narrow_resonance_scalar_tensor} confirm the expectation that the contributions from heavy scalars are small. Similarly, the net effect from tensor resonances remains moderate, but we observe a cancellation between the scalar functions $\bar\Pi_{1,2}$ and $\bar\Pi_{3\text{--}12}$, which, in the OPE limit, correspond to the longitudinal and transverse form factors $w_L$ and $w_T$, respectively. In view of the rather drastic approximation~\eqref{TFF_tensor}, an improved calculation of tensor-meson contributions, particularly of the $f_2(1270)$ as a predominantly elastic $\pi\pi$ resonance, is well motivated~\cite{Ludtke:2023hvz}, see also Sec.~\ref{sec:amu} for a discussion of possible avenues for future improvements.

\begin{table}[t]
	\centering
	\renewcommand{\arraystretch}{1.3}
	\begin{tabular}{lrrrrr}
	\toprule
	Resonance & \multirow{3}{*}{$\sm \ [\text{GeV}^2]$} & \multirow{3}{*}{$f_1\to \phi\gamma$} & \multirow{3}{*}{$a_\mu[\bar\Pi_{1,2}]\ [10^{-11}]$} & \multirow{3}{*}{$a_\mu[\bar\Pi_{3\text{--}12}]\ [10^{-11}]$}
	& \multirow{3}{*}{Sum $[10^{-11}]$}\\
	$\mA \ [\text{GeV}]$ & & & & &\\
	$\widetilde{\Gamma}_{\gamma\gamma}\ [\text{keV}]$ & & & & &\\
	\midrule
	 $f_1(1285)$&
	\multirow{2}{*}{$1.5$}
	&
	No
	& $2.9(7)$ & $2.1(5)$ & $5.0(1.1)$\\
	\multirow{2}{*}{$1.2818(5)$} &  &
	Yes
	& $1.9(5)$ & $1.3(4)$ & $3.1(9)$\\
	 &
	\multirow{2}{*}{$2.0$}
	&
	No
	& $2.2(5)$ & $1.6(4)$ & $3.8(1.0)$\\
	$3.5(8)$&  &
	Yes
	 & $1.4(5)$ & $1.0(4)$ & $2.4(8)$\\\midrule
	 $f_1'(1420)$ &
	\multirow{2}{*}{$1.5$}
	&
	No
	& $2.0(5)$ & $1.4(4)$ & $3.4(9)$\\
	\multirow{2}{*}{$1.4284(15)$} &  &
	Yes
	& $1.2(4)$ & $0.8(3)$ & $2.0(7)$\\
	 &
	\multirow{2}{*}{$2.0$}
	&
	No
	& $1.5(4)$ & $1.1(3)$ & $2.6(6)$\\
	$3.2(9)$ &  &
	Yes
	& $0.9(3)$ & $0.7(2)$ & $1.6(5)$\\\midrule
	$a_1(1260)$ &
	\multirow{2}{*}{$1.5$}
	&
	No
	& $2.2(4)$ & $1.5(3)$ & $3.8(7)$\\
	\multirow{2}{*}{$1.23(4)$} &  &
	Yes
	& $1.3(4)$ & $0.9(3)$ & $2.2(6)$\\
	 &
	\multirow{2}{*}{$2.0$}
	&
	No
	& $1.8(4)$ & $1.3(3)$ & $3.1(6)$\\
	$2.0(4)$ &  &
	Yes
	& $1.1(3)$ & $0.8(3)$ & $1.9(6)$\\\midrule
	 \multirow{4}{*}{Sum}&
	\multirow{2}{*}{$1.5$}
	&
	No
	& $7.2(1.4)$ & $5.0(1.0)$ & $12.2(2.3)$\\
	 &  &
	Yes
	& $4.4(1.2)$ & $3.0(9)$ & $7.3(2.0)$\\
	&
	\multirow{2}{*}{$2.0$}
	&
	No
	& $5.5(1.1)$ & $4.0(8)$ & $9.5(1.9)$\\
	&  &
	Yes
	& $3.4(1.0)$ & $2.5(8)$ & $5.9(1.7)$\\
\bottomrule
	\renewcommand{\arraystretch}{1.0}
	\end{tabular}
	\caption{Axial-vector narrow-resonance contributions for $Q_0=1.5\GeV$. The uncertainties are propagated from the parameterizations provided in Ref.~\cite{Hoferichter:2023tgp}, including all correlations among $\{C_s,C_{a_1},C_{a_2},\theta_A\}$ (the correlation coefficient of $C_s$ with $\theta_A$ is $\simeq -0.33$, while the correlations between $C_{a_{1/2}}$ and $\theta_A$ can be neglected). The value for $\widetilde{\Gamma}_{\gamma\gamma}^{a_1}$ assumes $U(3)$ symmetry, as do the TFF parameterizations for $f_1'$ and $a_1$.}
	\label{tab:narrow_resonance_axial}
\end{table}

The most important hadronic contributions not evaluated dispersively so far concern axial-vector states, see Table~\ref{tab:narrow_resonance_axial}. In this case, we use the dedicated work of Refs.~\cite{Zanke:2021wiq,Hoferichter:2023tgp} to determine the three TFFs of the $f_1(1285)$ from a global data analysis. At present, three couplings $\{C_s,C_{a_1},C_{a_2}\}$ in a VMD-inspired parameterization, corresponding  to the normalizations of the symmetric and the two antisymmetric TFFs, can be extracted, as well as the mixing angle $\theta_A$ with the $f_1'(1420)$, using data from $e^+e^-\to e^+e^- A$, $A=f_1,f_1'$~\cite{Achard:2001uu,Achard:2007hm}, $e^+e^-\to f_1\pi^+\pi^-$~\cite{BaBar:2007qju,BaBar:2022ahi}, and radiative decays~\cite{Zanke:2021wiq,Hoferichter:2023tgp,ParticleDataGroup:2024cfk}. In this context, it was observed that including a constraint from $f_1\to \phi\gamma$, related by flavor symmetry, led to increased tensions in the global fit, possibly reflecting the rather tenuous data situation. In Table~\ref{tab:narrow_resonance_axial} we show the results if the constraint from $f_1\to\phi\gamma$ is included, but since the matching to SDCs for the HLbL tensor clearly favors the fit without, we will use the latter variant for the further analysis. For a complete description, the asymptotic contributions to the TFFs~\cite{Hoferichter:2020lap} need to be added, in
App.~\ref{app:mass_effects} we detail the precise form we choose to both  incorporate effects of the axial-vector mass~\cite{Zanke:2021wiq} and ensure that the low-energy VMD form remains unaffected. These additional contributions depend on an effective decay constant $F_A^\text{eff}$, which can be estimated from light-cone sum rules~\cite{Yang:2007zt,Hoferichter:2020lap}, and a parameter $\sm$ that governs the transition to the asymptotic region.

\begin{figure}[t]
	\centering
	\includegraphics[width=0.9\linewidth]{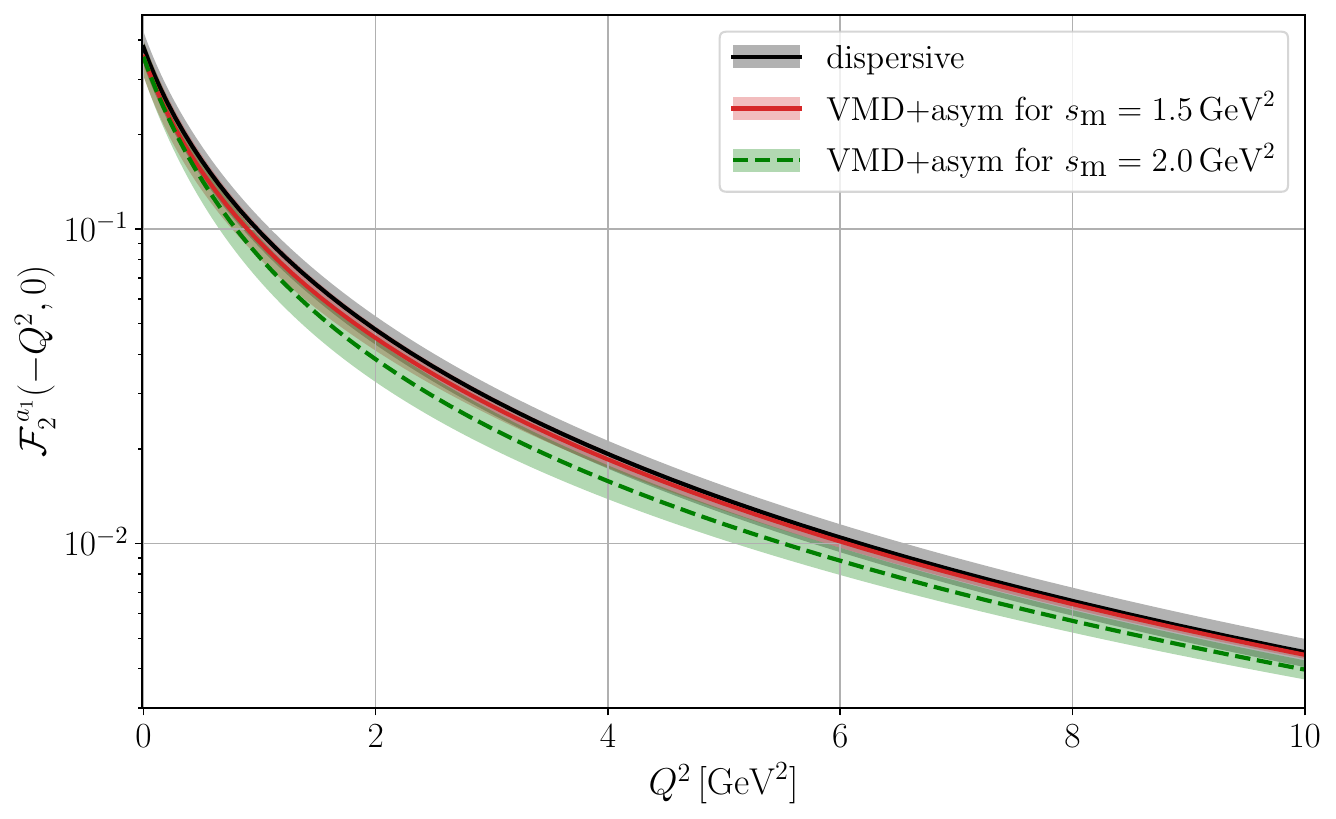}
	\caption{Singly-virtual TFF for the $a_1$, comparing the VMD parameterization from Ref.~\cite{Hoferichter:2023tgp}, supplemented by the asymptotic contribution in the formulation from App.~\ref{app:mass_effects} for two values of $\sm$, to the dispersive analysis of Ref.~\cite{Ludtke:2024ase}.}
	\label{fig:a1_singly_virtual_plot}
\end{figure}

Another cross check is provided by the dispersive calculation of the singly-virtual TFF for the $a_1$ from Ref.~\cite{Ludtke:2024ase}, see Fig.~\ref{fig:a1_singly_virtual_plot}. The calculation of the $a_1$ TFF from Ref.~\cite{Hoferichter:2023tgp} relies on $U(3)$ assumptions, but remarkably, the resulting normalization agrees perfectly with the dispersive analysis. Also the momentum dependence matches well, best for a value of $\sm=1.5\GeV^2$. In Table~\ref{tab:narrow_resonance_axial} we include results when $\sm$ is varied, but for the remainder of the analysis we will consider $\sm=1.5\GeV^2$ fixed, motivated by the comparison to the dispersive calculation of the singly-virtual $a_1$ TFF. Accordingly, we propagate the uncertainties from the fit without $f_1\to\phi\gamma$, including the significant correlations among $\{C_s,C_{a_1},C_{a_2},\theta_A\}$, in the contribution of the axial-vector resonances. In the end, we will also consider an additional systematic uncertainty of $30\%$ motivated by the required $U(3)$ assumptions, which then ensures consistency with most of the fit variants summarized in Table~\ref{tab:narrow_resonance_axial}.
These results also
show that the non-pole effects in $\bar\Pi_{1,2}$ are sizable, so that these axial-vector contributions differ quite substantially from their analogs in a dispersion relation in triangle kinematics, where non-pole terms instead arise from the intermediate states in the TFF~\cite{Ludtke:2023hvz,Ludtke:2024ase}. In the context of the effective poles discussed in App.~\ref{app:effective_poles}, we even observe that the contribution from $\bar\Pi_{3\text{--}12}$ can change sign between the two formulations, mainly due to the non-pole terms in $\hat\Pi_4$. Such a reshuffling of intermediate-state contributions between the two dispersive approaches is expected and has already been analyzed in detail for the case of the $VVA$ correlator~\cite{Ludtke:2024ase}.

\begin{table}[t]
	\centering
	\renewcommand{\arraystretch}{1.3}
	\begin{tabular}{lrrr}
	\toprule
	 & $a_\mu[\bar\Pi_{1,2}] \ [10^{-11}]$ & $a_\mu[\bar\Pi_{3\text{--}12}] \ [10^{-11}]$ & Sum $[10^{-11}]$\\\midrule
LO & $5.9$ & $1.8$ & $7.7$\\
NLO $\times\frac{\pi}{\alpha_s(Q_0)}$ & $-5.5$ & $-2.1$ & $-7.6$\\
LO+NLO & $5.3^{+0.1}_{-0.2}$ & $1.6^{+0.0}_{-0.1}$ & $6.9^{+0.2}_{-0.3}$\\
\bottomrule
	\renewcommand{\arraystretch}{1.0}
	\end{tabular}
	\caption{Quark loop and $\alpha_s$ corrections for $Q_0=1.5\GeV$~\cite{Bijnens:2021jqo}. The scale in $\alpha_s(\mu)$ is varied within $[Q_0/\sqrt{2},Q_0\sqrt{2}]$, the decoupling scales $\mu_c=3\GeV$, $\mu_b=m_b$ by a factor $2$, leading to $\alpha_s(Q_0)=0.35^{+0.11}_{-0.06}$~\cite{Herren:2017osy,Chetyrkin:2000yt}.}
	\label{tab:quark_loop}
\end{table}

\section{Matching to short-distance constraints}
\label{sec:matching_SDCs}

\begin{figure}[t]
	\centering
	\includegraphics[width=0.88\linewidth]{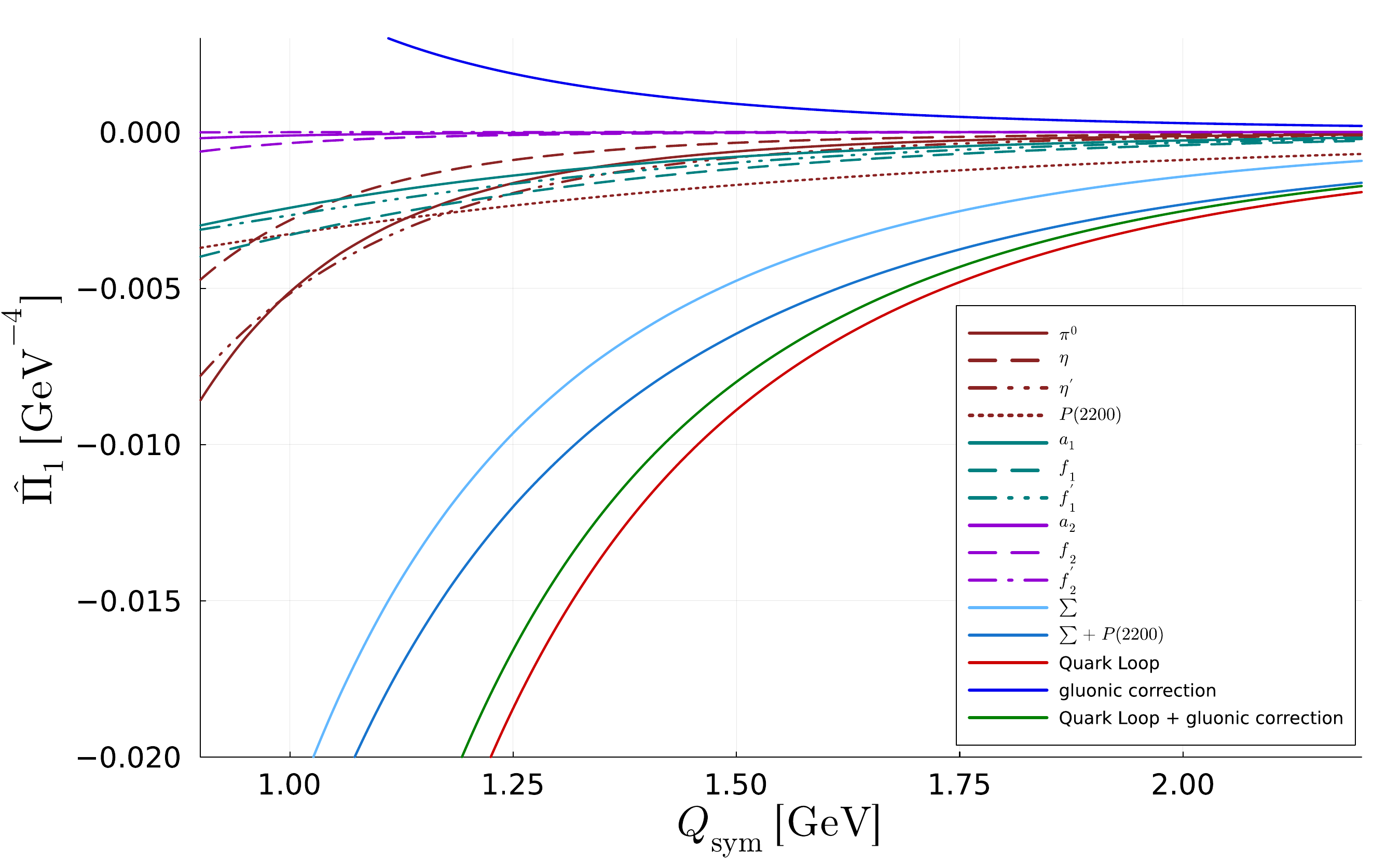}
	\caption{Matching of the hadronic contributions to $\hat \Pi_1$ to SDCs, along the line $Q_\text{sym}=Q_1=Q_2=Q_3$. The hadronic states are labeled as in the main text, $\sum=\pi^0+\eta+\eta'+a_1+f_1+f_1'+a_2+f_2+f_2'$, and $P(2200)$ denotes an effective pseudoscalar pole to help minimize the gap between the sum of hadronic states and pQCD. }
	\label{fig:hatPi1}
\end{figure}

\begin{figure}[t]
	\centering
	\includegraphics[width=0.86\linewidth]{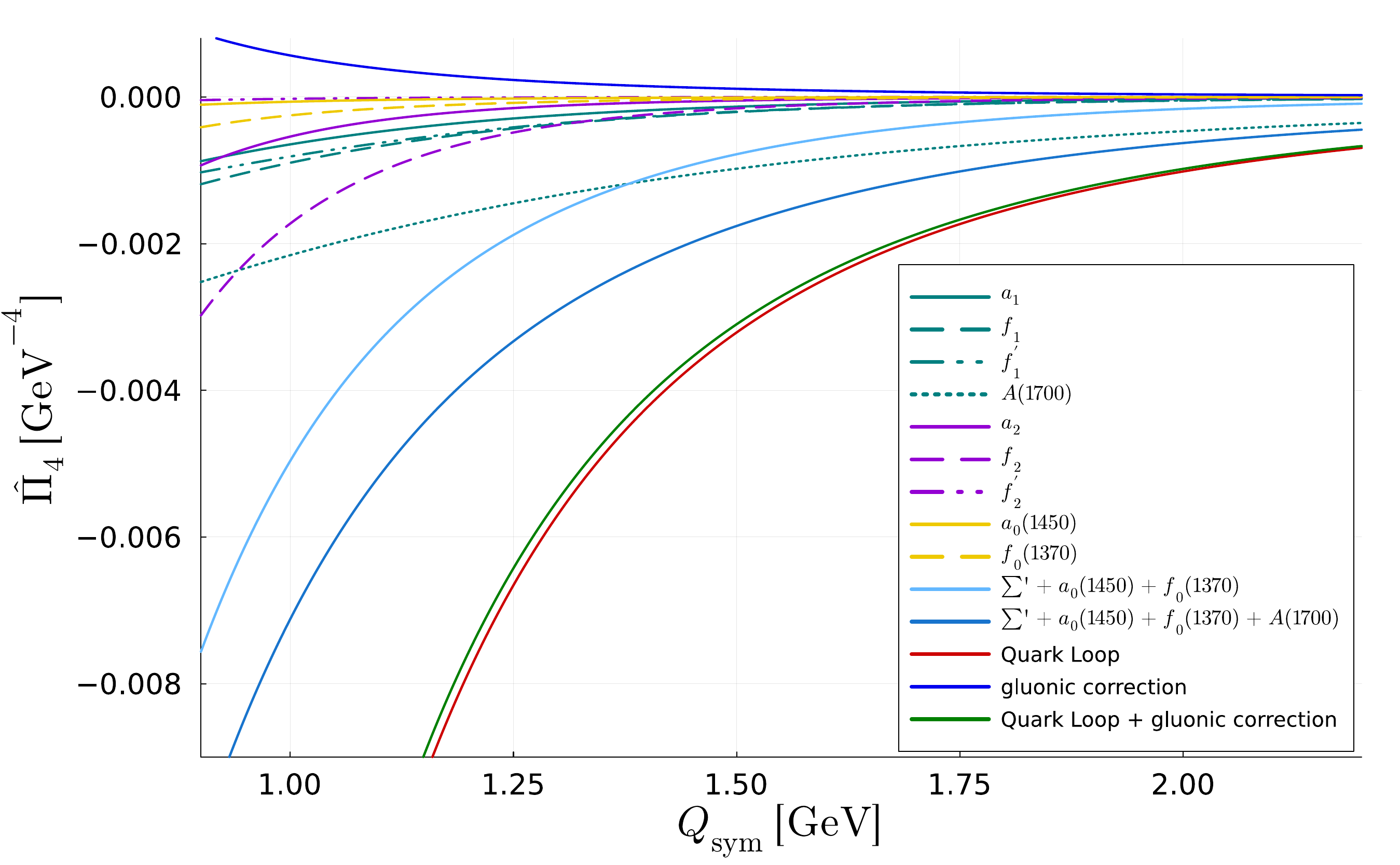}
	\includegraphics[width=0.86\linewidth]{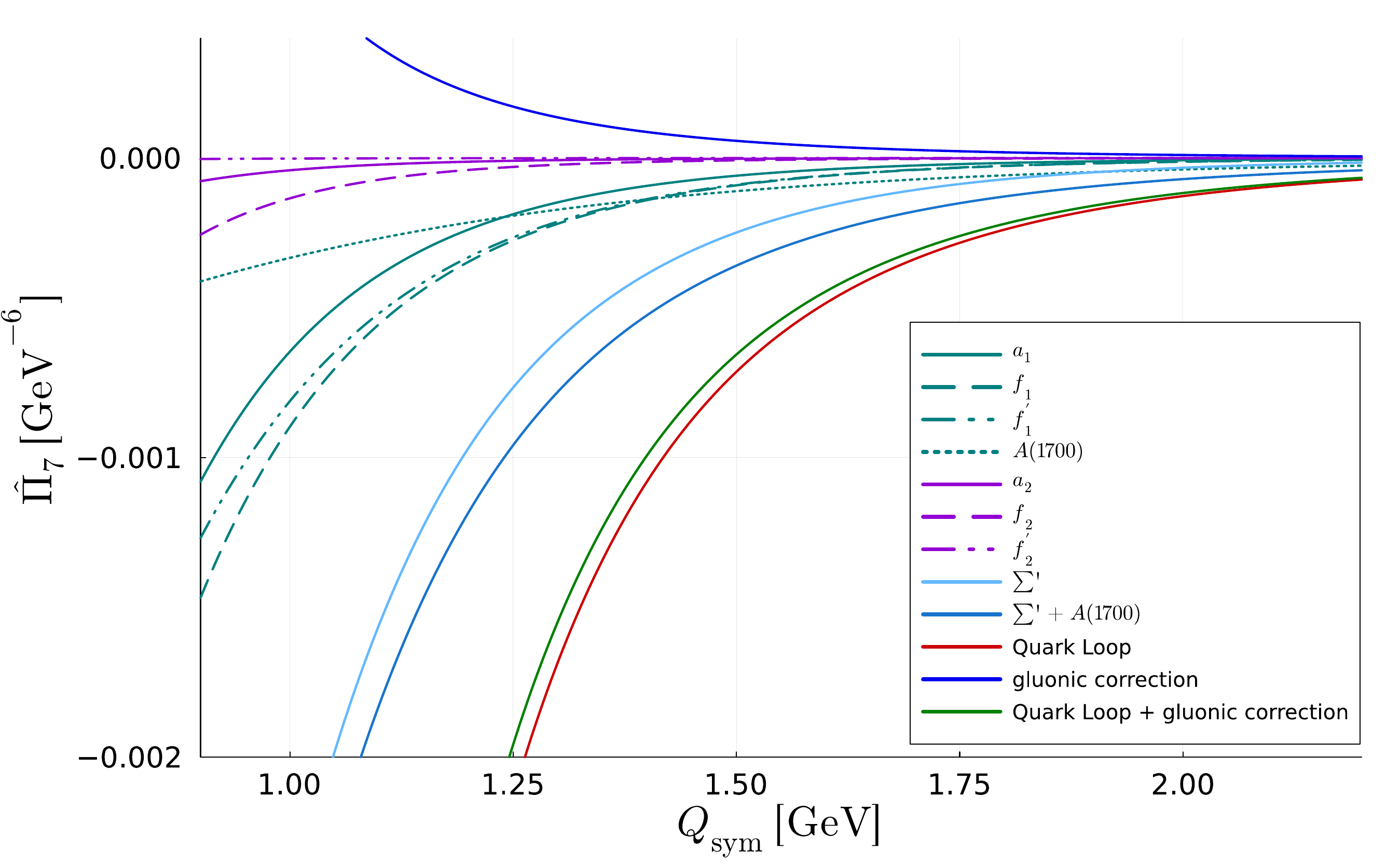}
	\caption{Analog of  Fig.~\ref{fig:hatPi1} for $\hat \Pi_{4,7}$, with axial-vector effective pole $A(1700)$ and $\sum'=a_1+f_1+f_1'+a_2+f_2+f_2'$.}
	\label{fig:hatPi47}
\end{figure}

Arguably the most important aspect regarding the evaluation of the subleading contributions to HLbL scattering concerns a smooth matching between the low-energy hadronic descriptions and SDCs. To this end, we vary the matching scale $Q_0$ between $Q_0=1.2\GeV$, the lowest scale at which $\alpha_s$ corrections in perturbative QCD (pQCD) can still be controlled, and $Q_0=2\GeV$, the highest scale at which a hadronic description only including the states discussed in Sec.~\ref{sec:narrow_resonance} still has a chance to be meaningful. This defines the purely hadronic region $Q_i\leq Q_0$ and the asymptotic region $Q_i\geq Q_0$, see Table~\ref{tab:quark_loop}.

\begin{figure}[t]
	\centering
	\includegraphics[width=0.86\linewidth]{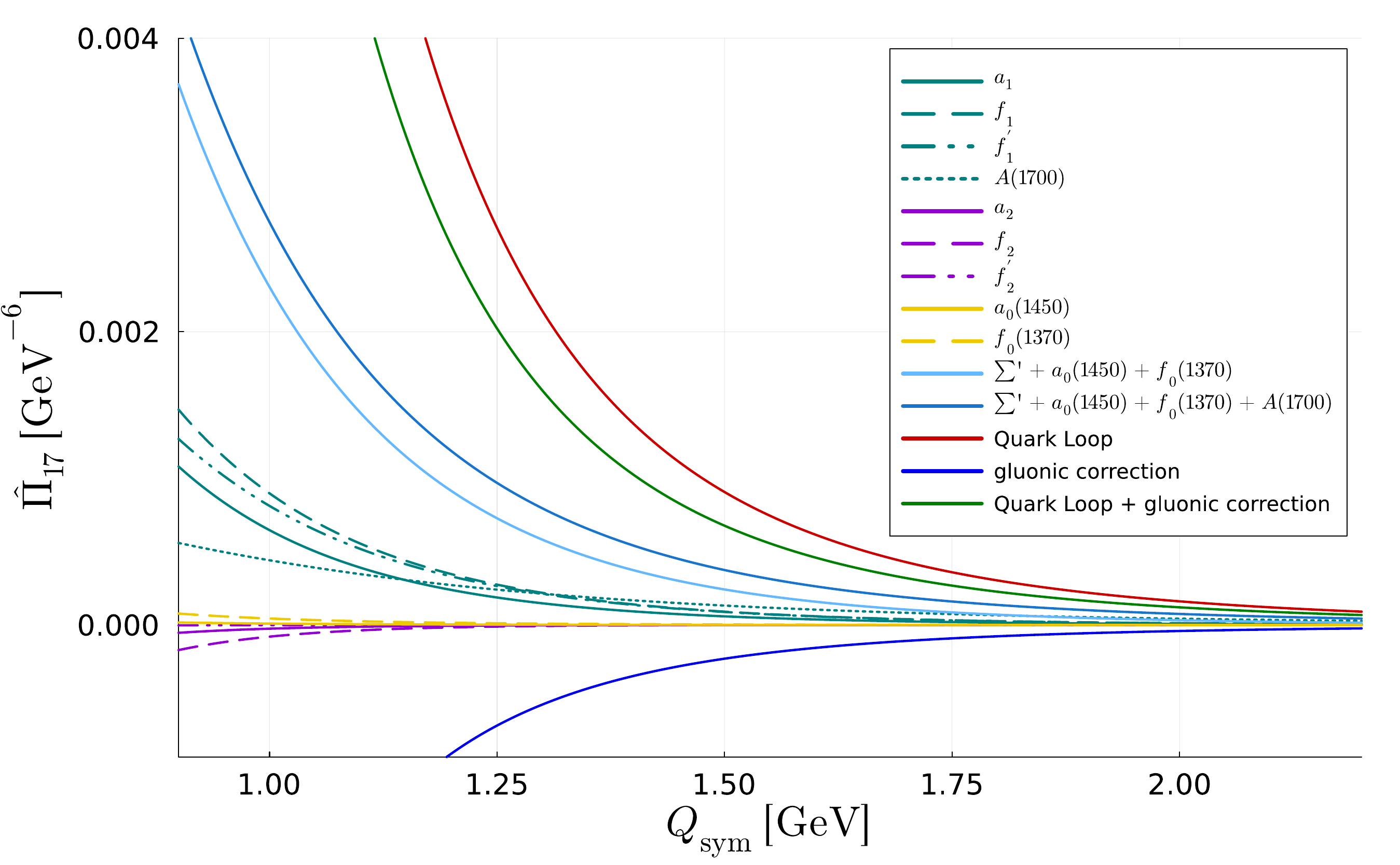}
	\includegraphics[width=0.86\linewidth]{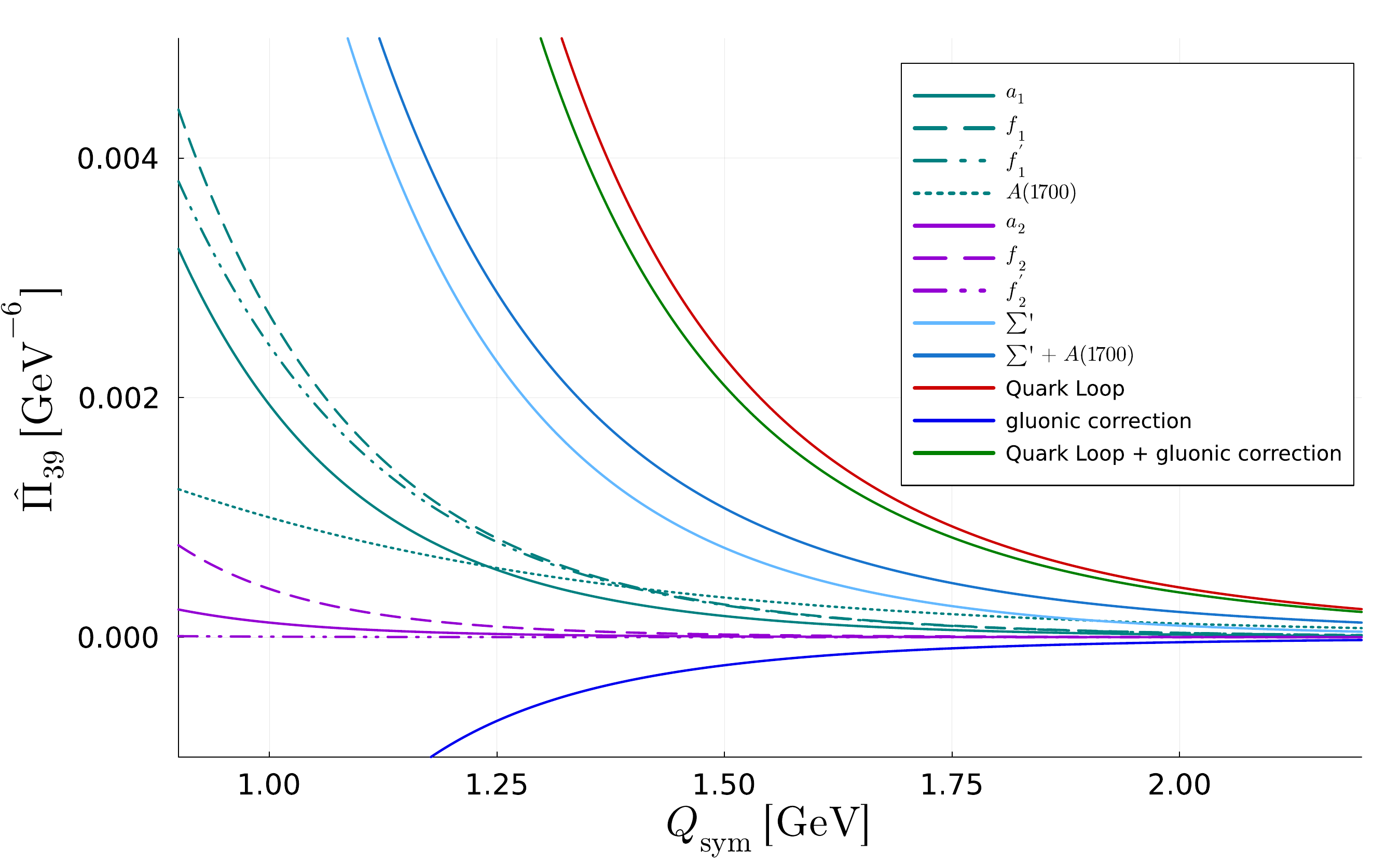}
	\caption{Analog of  Fig.~\ref{fig:hatPi47} for $\hat \Pi_{17,39}$.}
	\label{fig:hatPi1739}
\end{figure}

The matching between the two regions is illustrated along the line $Q_\text{sym}=Q_1=Q_2=Q_3$ in Figs.~\ref{fig:hatPi1}--\ref{fig:hatPi1739} ($\hat\Pi_{54}$ vanishes in this fully symmetric limit). One observes that the sum of hadronic states indeed reproduces a large fraction of the expected pQCD contribution, where the tails of the pseudoscalar poles from $\pi^0$, $\eta$, $\eta'$, the axial-vector contributions, and also the tensor states yield relevant effects, while the heavy scalars are largely negligible. With a finite set of hadronic states one cannot expect to  satisfy the SDCs exactly, so from this perspective the agreement appears rather decent.

However, one may still be concerned that the remaining states imply a low-energy effect that is not fully captured by the scale variation in $Q_0$ alone. For this reason, we also study the impact of adding an effective pole, constructed in such a way as to fulfill the SDCs in a hadronic realization, see App.~\ref{app:effective_poles}. The simplest way to achieve this proceeds via an effective pseudoscalar pole in triangle kinematics in $\hat\Pi_1$, and an axial-vector one for the other scalar functions. The overall couplings of the effective poles are determined such that the SDCs are fulfilled asymptotically by the sum of hadronic states and effective poles in the symmetric limit $Q_\text{sym}=Q_1=Q_2=Q_3$. We vary the scale in their TFFs in the same range as $Q_0$ and set the masses to $M_P^\text{eff}=2.2\GeV$, $M_A^\text{eff}=1.7\GeV$,\footnote{$M_A^\text{eff}=1.7\GeV$ is motivated by the first excited axial-vector resonance, the $a_1(1640)$, while $M_P^\text{eff}=2.2\GeV$ is the lowest mass for which the result does not exceed the pQCD curve at larger $Q_\text{sym}$, reflecting the scant evidence for excited-pseudoscalar two-photon couplings below $2\GeV$~\cite{Colangelo:2019uex,ParticleDataGroup:2024cfk}. In either case, the result is much more sensitive to the choice of scale in the TFF than the mass of the effective pole.} which leads to a visible improvement in the matching also at finite values of $Q_\text{sym}$, as shown in Figs.~\ref{fig:hatPi1}--\ref{fig:hatPi1739}.
We checked that the matching looks similarly reasonable in the asymmetric directions, but no  additional constraints that would merit implementation arise.
To estimate the potential impact of the remaining mismatch,
we will include the low-energy effects generated by the effective poles in our final error estimate.

Finally, we turn to the mixed region, in which no clear hierarchy between the $Q_i$ and $Q_0$ exists. In part of the parameter space, say $Q_3^2\ll \hat Q^2=(Q_1^2+Q_2^2)/2$, one can apply an OPE to relate the HLbL tensor to the $VVA$ correlator~\cite{Melnikov:2003xd,Vainshtein:2002nv,Knecht:2003xy}. In practice, this constraint proves particularly useful due to two recent developments: first, in Ref.~\cite{Bijnens:2024jgh} a remarkable cancellation at the level of the $a_\mu$ integration was found among non-perturbative form factors at higher orders in the OPE, rendering the leading expression in terms of the longitudinal and transverse form factors $w_{L,T}(q^2)$ exceptionally robust. Second, in Ref.~\cite{Ludtke:2024ase} a dedicated dispersive analysis of the $VVA$ correlator was performed, including phenomenological results for the triplet contributions to $w_{L,T}(q^2)$. In App.~\ref{app:VVA_matching} we detail the precise implementation at the level of the scalar functions $\hat\Pi_i$ and discuss the phenomenological input for octet and singlet components.

\begin{figure}[t]
	\centering
	\includegraphics[width=0.9\linewidth]{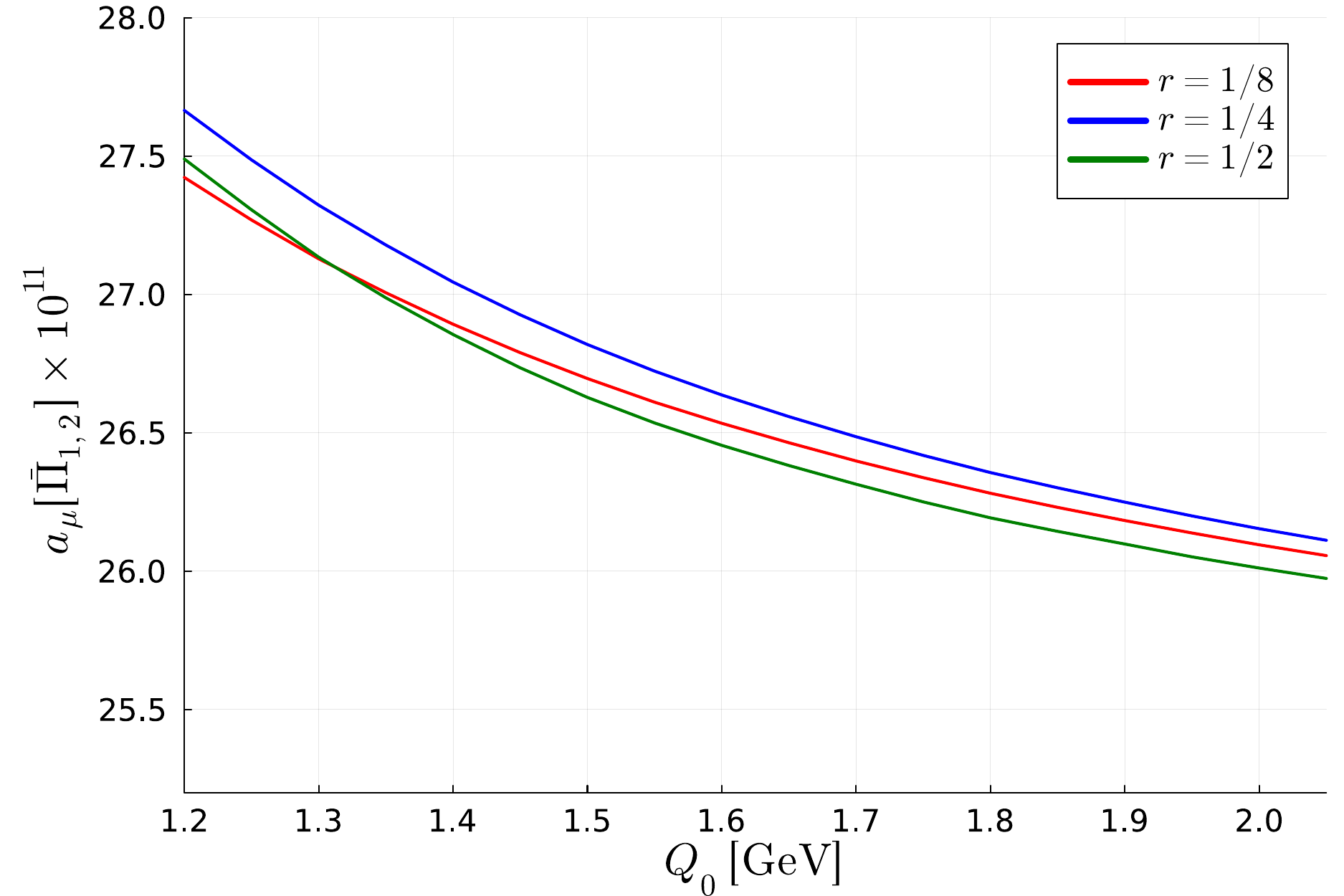}
	\includegraphics[width=0.9\linewidth]{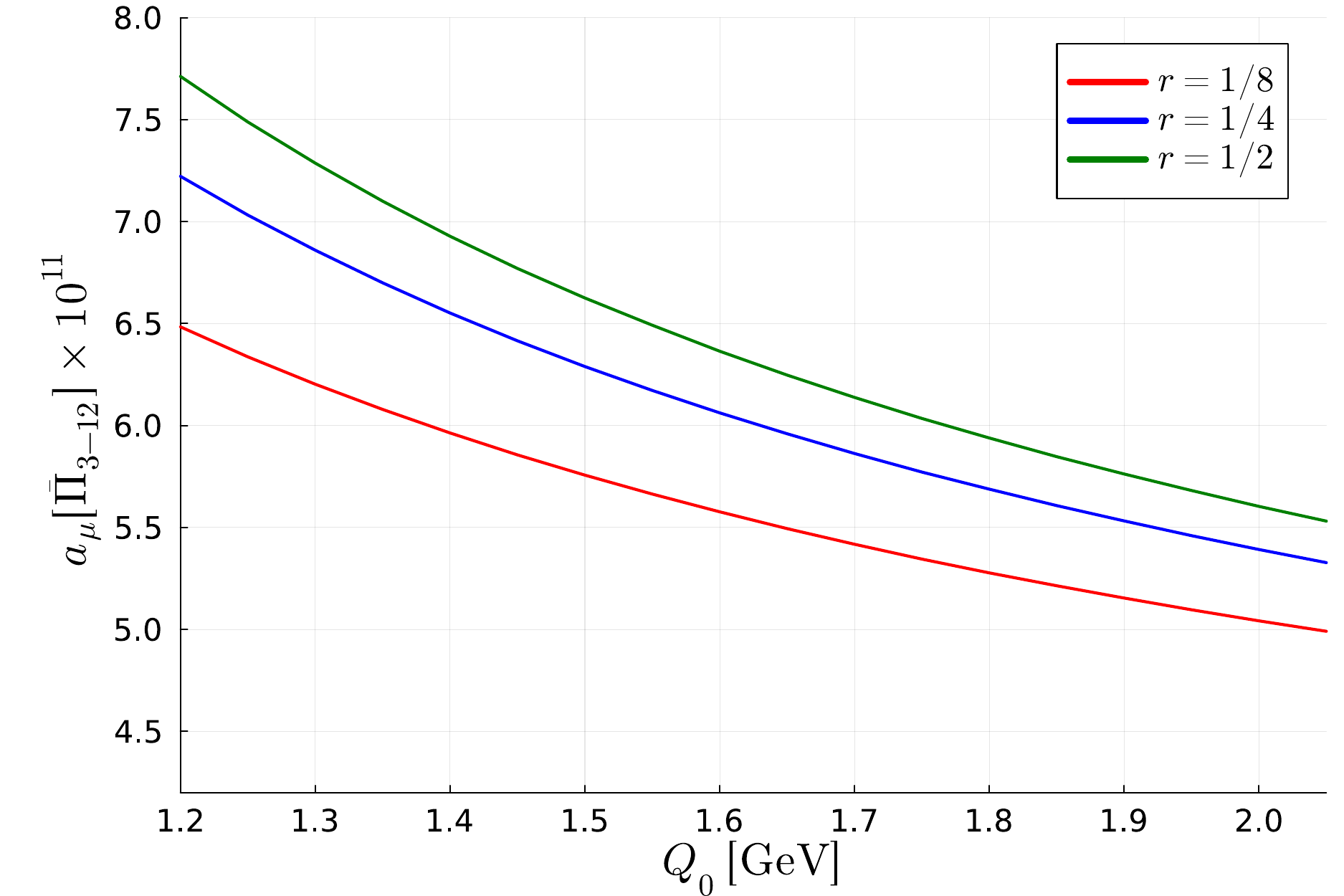}
	\caption{Dependence of $a_\mu[\bar\Pi_{1,2}]$ (upper) and   $a_\mu[\bar\Pi_{3\text{--}12}]$ (lower) on $Q_0$ and $r$.}
	\label{fig:amuLT}
\end{figure}

In the $a_\mu$ evaluation, we will use these OPE results whenever
\beq
\label{mixed_OPE}
Q_3^2\leq r \frac{Q_1^2+Q_2^2}{2},\qquad Q_1^2\geq Q_0^2,\qquad Q_2^2\geq Q_0^2,\qquad Q_{3}^{2}\leq Q_{0}^{2},
\eeq
and for the crossed versions with small $Q_1^2$ or $Q_2^2$. The parameter $r$, which controls the size of the corrections to the OPE, is set to $r=1/4$ and varied by a factor $2$ in either direction. For the remainder of the mixed region, in which the OPE condition~\eqref{mixed_OPE} does not apply, we use the hadronic description. In both the pQCD region and the OPE part of the mixed region we subtract the tails of $\pi^0$, $\eta$, $\eta'$, to be able to continue to use the full evaluations~\cite{Hoferichter:2018dmo,Hoferichter:2018kwz,Holz:2024lom,Holz:2024diw} of these pseudoscalar poles over the entire integration domain without incurring double counting.

\begin{figure}[t]
	\centering
	\includegraphics[width=0.9\linewidth]{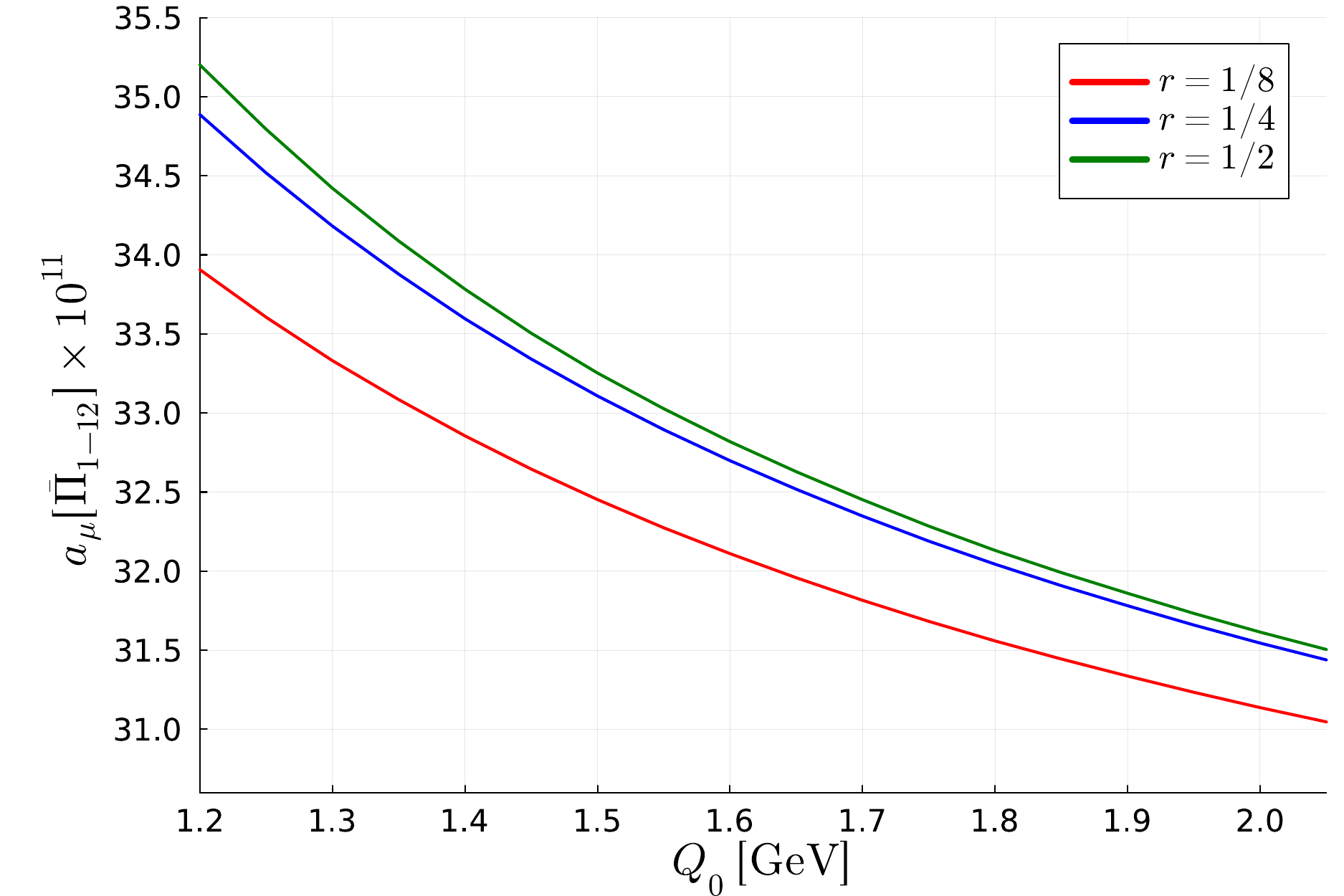}
	\caption{Dependence of the total $a_\mu[\bar\Pi_{1\text{--}12}]$  on $Q_0$ and $r$.}
	\label{fig:amutotal}
\end{figure}

\section{Contributions to $\boldsymbol{a_\mu}$}
\label{sec:amu}

As the final step in evaluating the subleading contributions to $a_\mu$, we study the dependence on $Q_0$ and $r$, see Figs.~\ref{fig:amuLT} and \ref{fig:amutotal}. These figures show that the variation remains very moderate, indicating good stability of the matching between hadronic contributions and SDCs. In total, we obtain the following results for the sum of the subleading contributions
\begin{align}
\label{result_subleading}
 a_\mu[\bar\Pi_{1,2}]&=26.9(2.1)_\text{exp}(1.0)_\text{match}(3.7)_\text{sys}(3.2)_\text{eff}[5.4]_\text{total}\times 10^{-11},\notag\\
 a_\mu[\bar\Pi_{3\text{--}12}]&=6.3(1.5)_\text{exp}(1.4)_\text{match}(0.2)_\text{sys}(2.2)_\text{eff}[3.0]_\text{total}\times 10^{-11},\notag\\
 a_\mu[\bar\Pi_{1\text{--}12}]&=33.2(3.3)_\text{exp}(2.2)_\text{match}(4.6)_\text{sys}(3.9)_\text{eff}[7.2]_\text{total}\times 10^{-11},
\end{align}
where the errors decompose as follows: (i) the first, experimental error is propagated from the two-photon decay widths of the heavy scalars and tensor mesons as well as the axial-vector TFFs; (ii) the second component reflects the uncertainty in the matching, defined as the maximal variation seen when varying $Q_0\in[1.2,2.0]\GeV$ and $r\in[1/8,1/2]$; (iii) the third error gives a $30\%$ uncertainty on the sum of  the hadronic contributions, to reflect the systematic uncertainties due to the use of $U(3)$ relations for the axial-vector TFFs and the simplified assumptions for the tensor TFFs;\footnote{For the total $a_\mu[\bar\Pi_{1\text{--}12}]$ value we add an additional $100\%$ uncertainty for the tensor contribution, to protect against the strong cancellation observed between $a_\mu[\bar\Pi_{1,2}]$ and $a_\mu[\bar\Pi_{3\text{--}12}]$ in this case, see Table~\ref{tab:narrow_resonance_scalar_tensor}.} (iv) the fourth error indicates the uncertainties in the effective-pole estimate, with the central value defined from the symmetric asymptotic matching at TFF scale $1.5\GeV$, and variations including the asymmetric matching and TFF scale within the same range as $Q_0$.\footnote{To avoid being sensitive to the cancellation between the pseudoscalar and axial-vector effective pole contributions, we determine the uncertainty for the total as the quadratic sum of the two effective-pole errors.} A more detailed breakdown of the various contributions is provided in Table~\ref{tab:subleading}.

\begin{table}[t]
	\centering
	\scalebox{0.745}{
	\renewcommand{\arraystretch}{1.3}
	\begin{tabular}{llrrr}
	\toprule
	 Region & & $a_\mu[\bar\Pi_{1,2}] \ [10^{-11}]$ & $a_\mu[\bar\Pi_{3\text{--}12}] \ [10^{-11}]$ & Sum $[10^{-11}]$\\\midrule
\multirow{4}{*}{$Q_i<Q_0$} & $A=f_1,f_1',a_1$ &  $7.2(1.4)_\text{exp}$ & $5.0(1.0)_\text{exp}$ & $12.2(2.3)_\text{exp}$\\
& $S=f_0(1370),a_0(1450)$ & -- & $-0.7(3)_\text{exp}$ & $-0.7(3)_\text{exp}$\\
& $T=f_2,a_2,f_2'$ & $2.6(3)_\text{exp}$ & $-5.1(7)_\text{exp}$ & $-2.5(3)_\text{exp}$\\
& Effective poles & $2.5$ & $-0.4$ & $2.0$\\\midrule
\multirow{3}{*}{Mixed} & $A,S,T$ & $2.5(7)_\text{exp}$ & $1.3(3)_\text{exp}$ & $3.8(1.0)_\text{exp}$\\
& OPE & $6.3$ & $4.7$ & $10.9$\\
&Effective poles & $1.1$ & $0.1$& $1.2$\\\midrule
$Q_i>Q_0$ & pQCD & $4.8^{+0.1}_{-0.2}$ & $1.6^{+0.0}_{-0.1}$ & $6.3^{+0.2}_{-0.3}$\\\midrule
Sum & & $26.9(2.1)_\text{exp}(3.7)_\text{sys}(3.2)_\text{eff}$ & $6.3(1.5)_\text{exp}(0.2)_\text{sys}(2.2)_\text{eff}$ & $33.2(3.3)_\text{exp}(4.6)_\text{sys}(3.9)_\text{eff}$\\
\bottomrule
	\renewcommand{\arraystretch}{1.0}
	\end{tabular}}
	\caption{Summary of the various subleading contributions considered in this work, at the matching scale $Q_0=1.5\GeV$ and with the OPE applied for $Q_3^2<r(Q_1^2+Q_2^2)/2$, $r={1/4}$. In the regions in which OPE and pQCD are used, the tails of the pseudoscalar poles are subtracted to avoid double counting. The effective poles are determined from the matching in the symmetric asymptotic limit, with TFF scale $1.5\GeV$. The errors shown in the main part of the table are propagated from the experimental and pQCD uncertainties only (accounting for correlations), while for the sum also the systematic and effective-pole errors as discussed in the main text are reproduced, including a $100\%$ uncertainty for the total tensor contribution. To obtain the final errors in Eq.~\eqref{result_subleading}, the matching uncertainty from the
	variation of $Q_0$, $r$ needs to be added.}
	\label{tab:subleading}
\end{table}

To obtain the full HLbL value, we combine our result for $a_\mu^\text{HLbL}\big|_\text{subleading}$ from Eq.~\eqref{result_subleading} with the previously evaluated contributions $a_\mu^\text{HLbL}\big|_\text{disp}$ given in Table~\ref{tab:disp_summary} and the small charm-loop correction $a_\mu^\text{HLbL}\big|_\text{charm}=3(1)\times 10^{-11}$~\cite{Colangelo:2019uex}, obtaining
\beq
\label{amu_total}
a_\mu^\text{HLbL}\big|_\text{total}=a_\mu^\text{HLbL}\big|_\text{disp}+a_\mu^\text{HLbL}\big|_\text{subleading}+a_\mu^\text{HLbL}\big|_\text{charm}=
101.9(7.9)\times 10^{-11}.
\eeq
Our result agrees well with the previous white-paper evaluation~\cite{Aoyama:2020ynm} and the lattice-QCD calculation by the Mainz group~\cite{Chao:2021tvp,Chao:2022xzg}, while suggesting a slightly lower value than Refs.~\cite{Blum:2023vlm,Fodor:2024jyn}.

The breakdown of the uncertainties in Eq.~\eqref{result_subleading} suggests avenues for future improvements. First, the experimental error and part of the systematic uncertainty can be corroborated or even reduced with additional experimental input especially for axial-vector TFFs. Second, the evaluation of the tensor contributions should be completed using the dispersive approach in triangle kinematics~\cite{Ludtke:2023hvz}, to avoid the simplifying assumptions on their TFFs and, in the case of the $f_2(1270)$, verify the narrow-resonance approximation against an evaluation  of $D$-wave $\pi\pi$ rescattering~\cite{Hoferichter:2019nlq,Danilkin:2019opj}. Finally, the interplay of the dispersive approaches in four-point and triangle kinematics should also allow for an improved understanding of the matching to SDCs and potentially missing contributions at low virtualities, to improve upon the effective-pole estimate used in this work.

\section{Conclusions}
\label{sec:summary}

We presented a complete evaluation of subleading contributions in the dispersive approach to HLbL scattering. To this end, we combined recent advances in the determination of axial-vector TFFs, SDCs, and the $VVA$ correlator to assess the contributions of exclusive hadronic states up to a matching scale $Q_0$, the matching to SDCs, and the impact of the mixed regions. Overall, we found that the sum of exclusive states matches reasonably well onto the SDCs, and estimated the impact of potentially missing contributions at low virtualities by introducing effective poles to reproduce the asymptotic behavior in the direction of either symmetric or asymmetric momentum configurations. Heavy scalar resonances prove largely negligible in the matching, while the tails of the light pseudoscalars as well as tensor resonances need to be considered, the latter being implemented here using simplified assumptions on their TFFs. A detailed error breakdown is provided in Eq.~\eqref{result_subleading} and Table~\ref{tab:subleading}, leading to the final result for the total HLbL contribution given in Eq.~\eqref{amu_total}.

\begin{table}[t]
	\centering
	\renewcommand{\arraystretch}{1.3}
	\begin{tabular}{lrr}
	\toprule
	 Contribution & $a_\mu [10^{-11}]$ & Reference\\\midrule
	 $\pi^0$, $\eta$, $\eta'$ poles & $91.2^{+2.9}_{-2.4}$ & \cite{Hoferichter:2018dmo,Hoferichter:2018kwz,Holz:2024lom,Holz:2024diw}\\
	 $\pi^\pm$ box & $-15.9(2)$ & \cite{Colangelo:2017qdm,Colangelo:2017fiz}\\
	 $K^\pm$ box & $-0.5(0)$ & \cite{Stamen:2022uqh}\\
	 $S$-wave rescattering & $-9.1(1.0)$ & \cite{Colangelo:2017qdm,Colangelo:2017fiz,Danilkin:2021icn,Deineka:2024mzt}\\\midrule
	 Sum & $65.7^{+3.1}_{-2.6}$ &\\
\bottomrule
	\renewcommand{\arraystretch}{1.0}
	\end{tabular}
	\caption{Summary of previously evaluated contributions in dispersion theory. The $S$-wave rescattering subsumes effects that correspond to the light scalar resonances $f_0(500)$, $f_0(980)$, and $a_0(980)$.}
	\label{tab:disp_summary}
\end{table}

While the precision of HLbL scattering has now reached the level mandated by the upcoming final result of the Fermilab experiment, there are several aspects of the calculation that should be addressed in future work, to corroborate and potentially improve the uncertainties. First, for the axial-vector and tensor contributions, we had to rely on certain assumptions due to a lack of experimental input data and limitations of the dispersive approach in four-point kinematics, which could be addressed by new measurements, possible at BESIII~\cite{Redmer:2024bva,BESIII:2020nme} and Belle II~\cite{Belle-II:2018jsg}, and a complementary dispersive approach in triangle kinematics~\cite{Ludtke:2023hvz}, respectively. Especially in combination with the established approach in four-point kinematics,
the latter should also allow for an improved understanding of the matching to SDCs and uncertainties related to the truncation of hadronic states, in analogy to the analysis performed for $VVA$~\cite{Ludtke:2024ase}, but at the present stage we are convinced that the evaluation in Eq.~\eqref{amu_total} represents the best dispersive HLbL evaluation possible, with a realistic and conservative estimate of the remaining uncertainties.

\acknowledgments
We thank Gilberto Colangelo, Bastian Kubis, and Massimiliano Procura for decade-long collaboration on many aspects of the work presented here. We further thank
Johan Bijnens,  Nils Hermansson-Truedsson,  Jan L\"udtke, Antonio Rodr\'iguez-S\'anchez, and Matthias Steinhauser for valuable discussions, and Johan Bijnens for sharing code for the $\alpha_s$ corrections to the quark loop~\cite{Bijnens:2021jqo}.
Financial support by the SNSF (Project Nos.\ PCEFP2\_181117, PCEFP2\_194272, and TMCG-2\_213690) is gratefully acknowledged. 

\appendix

\section{Mass effects in the asymptotic contribution to axial-vector TFFs}
\label{app:mass_effects}

Constraints on the asymptotic form of axial-vector TFFs were derived in Ref.~\cite{Hoferichter:2020lap} from the leading term in the light-cone expansion. The most compact way of representing the result proceeds in terms of a derivative of the pseudoscalar TFFs
\beq
\F_2^\text{asym}(q_1^2,q_2^2)=3F_A^\text{eff}\mA^3\frac{\partial}{\partial q_1^2}\bar\F_P^\text{asym}(q_1^2,q_2^2),
\eeq
where $F_A^\text{eff}$ denotes the axial-vector decay constant~\cite{Hoferichter:2020lap} and the starting point for the pseudoscalar TFF is\footnote{The bar indicates the unconventional normalization, to distinguish this auxiliary quantity from a physical pseudoscalar TFF.}
\beq
\bar\F_P^\text{asym}(q_1^2,q_2^2)=-\frac{1}{3}\int_0^1du\frac{\phi_P(u)}{u q_1^2+(1-u)q_2^2-u(1-u)\mA^2},\qquad \phi_P(u)=6u(1-u).
\eeq
In a dispersive representation, only the high-energy part in the integral should be kept. Moreover, if one requires that $\bar\F_P^\text{asym}$ should not contribute to singly-virtual limits, the spectral function in the massless limit eventually leads to the form~\cite{Hoferichter:2018dmo,Hoferichter:2018kwz}
\beq
\label{FP_asym_1}
\bar\F_P^\text{asym}(q_1^2,q_2^2)=\int_{\sm}^\infty dx\frac{q_1^2q_2^2}{(x-q_1^2)^2(x-q_2^2)^2},
\eeq
and thus
\beq
\F_2^\text{asym}(q_1^2,q_2^2)=3F_A^\text{eff}\mA^3\int_{\sm}^\infty dx\frac{q_2^2(x+q_1^2)}{(x-q_1^2)^3(x-q_2^2)^2}.
\eeq
By construction, this function vanishes at $q_2^2=0$, while the doubly-virtual limit
\beq
\lim_{q^2\to\infty} \big(q^2\big)^2\F_2^\text{asym}(q^2,q^2)=\frac{1}{2}F_A^\text{eff}\mA^3
\eeq
is fulfilled. However, contrary to the pion TFF in Refs.~\cite{Hoferichter:2018dmo,Hoferichter:2018kwz}, the low-energy part of the axial-vector TFFs from Refs.~\cite{Zanke:2021wiq,Hoferichter:2023tgp} does not, by itself, saturate the singly-virtual limit, rendering Eq.~\eqref{FP_asym_1} inapplicable in this form. As a remedy, it is easiest to keep a total derivative in the derivation that only contributes at subleading orders in the asymptotic expansion, leading to
%\begin{align}
%\label{F_asym_massless}
% \F_2^\text{asym}(q_1^2,q_2^2)
% &=\frac{3F_A^\text{eff}\mA^3}{1+\beta}\frac{\partial}{\partial q_1^2}\int_{\sm}^\infty dx
 %\Bigg\{\frac{d}{dx}\bigg[\frac{x (2q_1^2-x)}{2(x-q_1^2)^2(x-q_2^2)}\bigg]
% +\frac{(1+\beta)q_1^2q_2^2}{(x-q_1^2)^2(x-q_2^2)^2}\Bigg\}\notag\\
 %
% &=\frac{3F_A^\text{eff}\mA^3}{1+\beta}\int_{\sm}^\infty dx
% \Bigg\{\frac{d}{dx}\bigg[\frac{x q_1^2}{(x-q_1^2)^3(x-q_2^2)}\bigg]
% +(1+\beta)\frac{q_2^2(x+q_1^2)}{(x-q_1^2)^3(x-q_2^2)^2}\Bigg\}.
%\end{align}
\begin{align}
\label{F_asym_massless}
 \F_2^\text{asym}(q_1^2,q_2^2)
 &=\frac{3F_A^\text{eff}\mA^3}{1+\beta}\int_{\sm}^\infty dx
 \Bigg\{\frac{d}{dx}\bigg[\frac{x\big[x(x-q_1^2)q_2^2-\gamma(x-q_2^2)\big(q_1^2\big)^2\big]}{\gamma(x-q_1^2)^4(x-q_2^2)^2}\bigg]\notag\\
 &\qquad+(1+\beta)\frac{q_2^2(x+q_1^2)}{(x-q_1^2)^3(x-q_2^2)^2}\Bigg\}.
\end{align}
This variant fulfills
\beq
\label{asym_limits}
\lim_{q^2\to\infty} \big(q^2\big)^2\F_2^\text{asym}(q^2,q^2)=\frac{1}{2}F_A^\text{eff}\mA^3,\qquad
\lim_{q^2\to\infty} \big(q^2\big)^2\F_2^\text{asym}(q^2,0)=\frac{3F_A^\text{eff}\mA^3}{1+\beta},
\eeq
and thus, in the limit $\beta=0$, respects both singly- and doubly-virtual asymptotic limits. In practice, the choice of a finite value of $\beta$ allows one to correct the singly-virtual coefficient implied by the low-energy part of the TFF. Moreover, Eq.~\eqref{F_asym_massless} does not contribute to the singly-virtual slope of the form factor, i.e.,
\beq
\label{asym_slope}
\F_2^\text{asym}(q_1^2,0)=
\Order\Big(\big(q_1^2\big)^2\Big).
\eeq
The properties~\eqref{asym_limits} and~\eqref{asym_slope} hold true for all values of the parameter $\gamma$, which can thus be used to adjust a smooth behavior for small virtualities (we use $\gamma=1.5$).

The spectral function for finite axial-vector masses was studied in Ref.~\cite{Zanke:2021wiq}. Using these results, we generalize Eq.~\eqref{F_asym_massless} according to
\begin{align}
 \F_2^\text{asym}(q_1^2,q_2^2)
 &=\frac{3F_A^\text{eff}\mA^3}{1+\beta}\int_{\sm}^\infty dx\Bigg\{\frac{d}{dx}\Bigg[\frac{4}{\mA^4}\bigg(1+\frac{\frac{2(x-q_1^2)^2-\mA^2 x}{W_1}+\frac{2(x-q_2^2)^2-\mA^2 x}{W_2}}{q_1^2+q_2^2-2x}\bigg)\notag\\
 %&\qquad\times\frac{2x q_1^2(x-q_2^2)^2}{\big(q_1^2q_2^2\big)^2-2xq_1^2q_2^2(q_1^2+q_2^2)+5x^2q_1^2q_2^2-x^3(q_1^2+q_2^2)}\bigg]
 &\qquad\times\frac{2x(x-q_2^2)\big[x(x-q_1^2)q_2^2-\gamma(x-q_2^2)\big(q_1^2\big)^2\big]}{\gamma(x-q_1^2)\Big[\big(q_1^2q_2^2\big)^2-2xq_1^2q_2^2(q_1^2+q_2^2)+5x^2q_1^2q_2^2-x^3(q_1^2+q_2^2)\Big]}\Bigg]
 \notag\\
 &-\frac{4(1+\beta)}{\mA^4}\frac{\partial}{\partial q_1^2}\bigg[\frac{q_2^2}{2x-q_1^2}\bigg(\frac{1}{2x-q_1^2-q_2^2}-\frac{1}{q_1^2-q_2^2}\bigg)\notag\\
 &\qquad\times\bigg(\frac{2(x-q_1^2)^2-\mA^2 x}{W_1}+q_1^2-x\bigg)+(q_1\leftrightarrow q_2)\bigg]\Bigg\},\notag\\
 W_i&=\sqrt{4(x-q_i^2)^2-4\mA^2 x+\mA^4},
\end{align}
constructed in such a way that the limit $\mA\to 0$ reproduces Eq.~\eqref{F_asym_massless}, while maintaining the low- and high-energy limits~\eqref{asym_limits} and~\eqref{asym_slope}.

\section{Matching to $\boldsymbol{VVA}$}
\label{app:VVA_matching}

In the limit in which two virtualities are much bigger than the third one, the HLbL tensor can be related to the $VVA$ correlator by OPE arguments~\cite{Melnikov:2003xd,Bijnens:2022itw,Bijnens:2024jgh}. The leading non-trivial constraints can be expressed as~\cite{Colangelo:2019uex}
\begin{align}
\label{hatPi_wL_wT}
 \hat{\Pi}_1&=-\frac{1}{\pi^2\hat q^2}\sum_{a=0,3,8}C_a^2w_L^{(a)}(q_3^2),\notag\\
 \hat{\Pi}_5&=\hat{\Pi}_6=-\hat q^2 \hat{\Pi}_{10}=-\hat q^2 \hat{\Pi}_{14}=\hat q^2 \hat{\Pi}_{17}=\hat q^2 \hat{\Pi}_{39}=2\hat q^2 \hat{\Pi}_{50}=2\hat q^2 \hat{\Pi}_{51}\notag\\
 &=-\frac{2}{3\pi^2\hat q^2}\sum_{a=0,3,8}C_a^2w_T^{(a)}(q_3^2),
\end{align}
where $\hat q = (q_1-q_2)/2$ with $|\hat q^2| \gg |q_3^2|$,
\beq
C_3=\frac{1}{6},\qquad C_8=\frac{1}{6\sqrt{3}},\qquad C_0=\frac{2}{3\sqrt{6}},
\eeq
and the other configurations follow from crossing symmetry. The relation for $\hat \Pi_{17}$ follows from $D=4$ OPE considerations~\cite{Bijnens:2022itw}, and by the same reasoning we put some of the ambiguities present in the matching from Ref.~\cite{Colangelo:2019uex} to zero
\beq
c_5^{(2)}=c_6^{(2)}=c_7^{(5)}=c_9^{(3)}=0.
\eeq
We emphasize that such a strict identification only works for asymptotically large $q_3^2$. At finite $q_3^2$, the ambiguities $c_i^{(n)}$ receive contributions both from the leading and higher-dimensional operators involving new non-perturbative form factors~\cite{Adam:2023}, see below for the example of axial-vector contributions to the $VVA$ correlator. However, these contributions to the ambiguities $c_i^{(n)}$ do not only cancel asymptotically, but the calculation of Ref.~\cite{Bijnens:2024jgh} shows that such effects even drop out at the level of the $a_\mu$ integration for finite $q_3^2$, up to even higher orders in the OPE.

The axial anomaly determines the non-singlet normalizations in the chiral limit according to~\cite{Adler:1969gk,Bell:1969ts}
\beq
\label{wLT_anomaly}
w_L^{(a)}(q^2)=2w_T^{(a)}(q^2)=\frac{6}{q^2}.
\eeq
In the longitudinal case, these constraints are already approximately satisfied by a simple model for the pseudoscalar poles\footnote{This expression follows in triangle kinematics when taking $M_P^2\to 0$ in the singly-virtual TFF.}
\beq
 \hat \Pi_1=\sum_{P=\pi^0,\eta,\eta'} \frac{F_{P\gamma\gamma}\bar F_\text{asym}^P}{3\hat q^2(M_P^2-q_3^2)},
\eeq
i.e., numerically, the corresponding functions
\begin{align}
\label{wL_mixed}
w_L^{(3)}(q^2)&=\frac{12\pi^2 F_{\pi\gamma\gamma}\bar F^{\pi}_\text{asym}}{q^2-\mpi^2},\notag\\
w_L^{(0,8)}(q^2)\equiv\frac{C_8^2w_L^{(8)}(q^2)+C_0^2w_L^{(0)}(q^2)}{C_8^2+C_0^2}&=\sum_{P=\eta,\eta'}\frac{4\pi^2 F_{P\gamma\gamma}\bar F^{P}_\text{asym}}{q^2-M_P^2},
\end{align}
fulfill the  normalizations~\eqref{wLT_anomaly} exactly for the pion due to
$2\pi^2F_{\pi\gamma\gamma}\bar F^{\pi}_\text{asym}=4\pi^2 F_\pi F_{\pi\gamma\gamma}=1$, and still approximately for $\eta$, $\eta'$
\beq
\frac{2\pi^2}{3}\big(F_{\eta\gamma\gamma}\bar F^{\eta}_\text{asym}+F_{\eta'\gamma\gamma}\bar F^{\eta'}_\text{asym}\big)\simeq 0.95,
\eeq
using normalizations from Ref.~\cite{ParticleDataGroup:2024cfk} and $\bar F^{\eta}_\text{asym}\simeq 186\MeV$, $\bar F^{\eta'}_\text{asym}\simeq 271\MeV$~\cite{Holz:2024lom}. Accordingly, Eq.~\eqref{wL_mixed} defines a plausible description in the mixed regions for $a=0,8$.

For the transverse component, one could similarly try to saturate the expected behavior~\eqref{hatPi_wL_wT} with axial-vector states. Matching the HLbL expressions for narrow axial-vector states to Eq.~\eqref{hatPi_wL_wT} and neglecting the ambiguities $c_i^{(n)}$, one obtains
\begin{align}
\label{wT_axial}
 w_T^{(3)}(q^2)&=\frac{27\pi^2 \frac{F_{a_1}^\text{eff}}{\Ma}\F_2^{a_1}(0,0)}{q^2-\Ma^2},\notag\\
 w_T^{(0,8)}(q^2)\equiv\frac{C_8^2w_T^{(8)}(q^2)+C_0^2w_T^{(0)}(q^2)}{C_8^2+C_0^2}&=\sum_{A=f_1,f_1'}\frac{9\pi^2 \frac{F_{A}^\text{eff}}{M_{A}}\F_2^{A}(0,0)}{q^2-M_{A}^2},
\end{align}
where we used the asymptotic constraints from Ref.~\cite{Hoferichter:2020lap}. This identification correctly reproduces all scalar functions apart from $\hat\Pi_{50,51}$, which are overestimated by a factor 2. In addition,
the matching implied by Eq.~\eqref{wT_axial} disagrees with a direct evaluation of the $VVA$ matrix element using axial-vector states~\cite{Ludtke:2024ase}, in which case one obtains a result larger by a factor $4/3$~\cite{Adam:2023}. Both mismatches are explained by
 non-vanishing contributions from $c_5^{(2)}$, $c_6^{(2)}$, which need to be canceled for large $q_3^2$ to satisfy the perturbative limit. These additional contributions to $c_5^{(2)}$, $c_6^{(2)}$ provide an example for the non-perturbative effects mentioned above.
In either way,
while the parametric dependence of Eq.~\eqref{wT_axial} matches Eq.~\eqref{wLT_anomaly},
the coefficients~\cite{Hoferichter:2020lap,Hoferichter:2023tgp}
\beq
27\pi^2 \frac{F_{a_1}^\text{eff}}{\Ma}\F_2^{a_1}(0,0)\simeq 8.6,\qquad
9\pi^2\sum_{A=f_1,f_1'}\frac{F_{A}^\text{eff}}{M_{A}}\F_2^{A}(0,0)\simeq 7.7,
\eeq
come out too large by almost a factor $3$, so this simple approach proves insufficient.

For this reason, we rely on the dispersive calculation from Ref.~\cite{Ludtke:2024ase} for the triplet components. To obtain a similar description for $a=0,8$, we first construct a simplified version for $a=3$. To this end, we combine ChPT constraints for the normalizations~\cite{Knecht:2020xyr,Masjuan:2020jsf} with OPE limits:
\begin{align}
 w_L^{(3)}(0)&=-\frac{24\pi^2(1+a_\pi)F_{\pi\gamma\gamma}F_\pi}{\mpi^2},\qquad
 w_T^{(3)}(0)=-\frac{24\pi^2a_\pi F_{\pi\gamma\gamma}F_\pi}{\mpi^2},\notag\\
 w_L^{(3)}(q^2)\Big|_\text{OPE}&=2w_T^{(3)}(q^2)\Big|_\text{OPE}=\frac{6}{q^2}\bigg[1+\frac{\kappa_\text{OPE}^{(3)}}{q^2}+\Order\Big(\big(q^2\big)^{-2}\Big)\bigg],
\end{align}
where $a_\pi=31.5(9)\times 10^{-3}$~\cite{Hoferichter:2018kwz} is the slope of the pion TFF and the OPE correction
\beq
\kappa_\text{OPE}^{(3)}
=\frac{8\pi^2\hat m\hat X}{3e}\simeq 0.012\GeV^2
\eeq
is expressed in terms of the tensor coefficients~\cite{Bali:2012jv,Bijnens:2019ghy,Ludtke:2024ase}
\beq
X_u=40.7(1.3)\MeV,\qquad X_d=39.4(1.4)\MeV,\qquad
X_s=53.0(7.2)\MeV,
\eeq
given at an $\overline{\text{MS}}$ scale $\mu=2\GeV$. Isospin-breaking corrections beyond $\hat m=(m_u+m_d)/2$, $\hat X=(X_u+X_d)/2$ have been neglected and for the quark masses we use $\hat m=3.49(7)\MeV$, $m_s=93.5(8)\MeV$~\cite{ParticleDataGroup:2024cfk,BMW:2010ucx,RBC:2012cbl,FermilabLattice:2018est,Lytle:2018evc,Bruno:2019vup,ExtendedTwistedMass:2021gbo}.

\begin{figure}[t]
	\centering
	\includegraphics[width=0.9\linewidth]{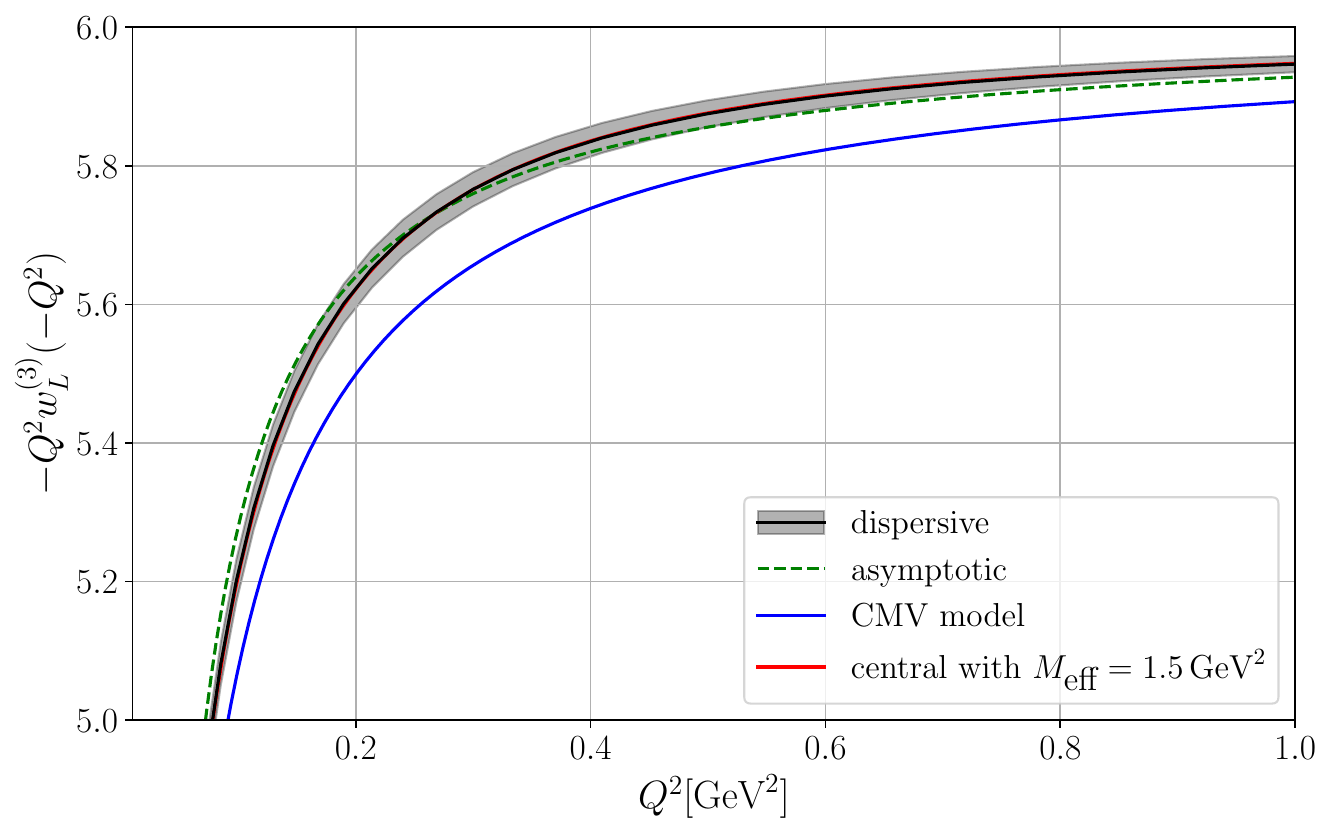}
	\includegraphics[width=0.9\linewidth]{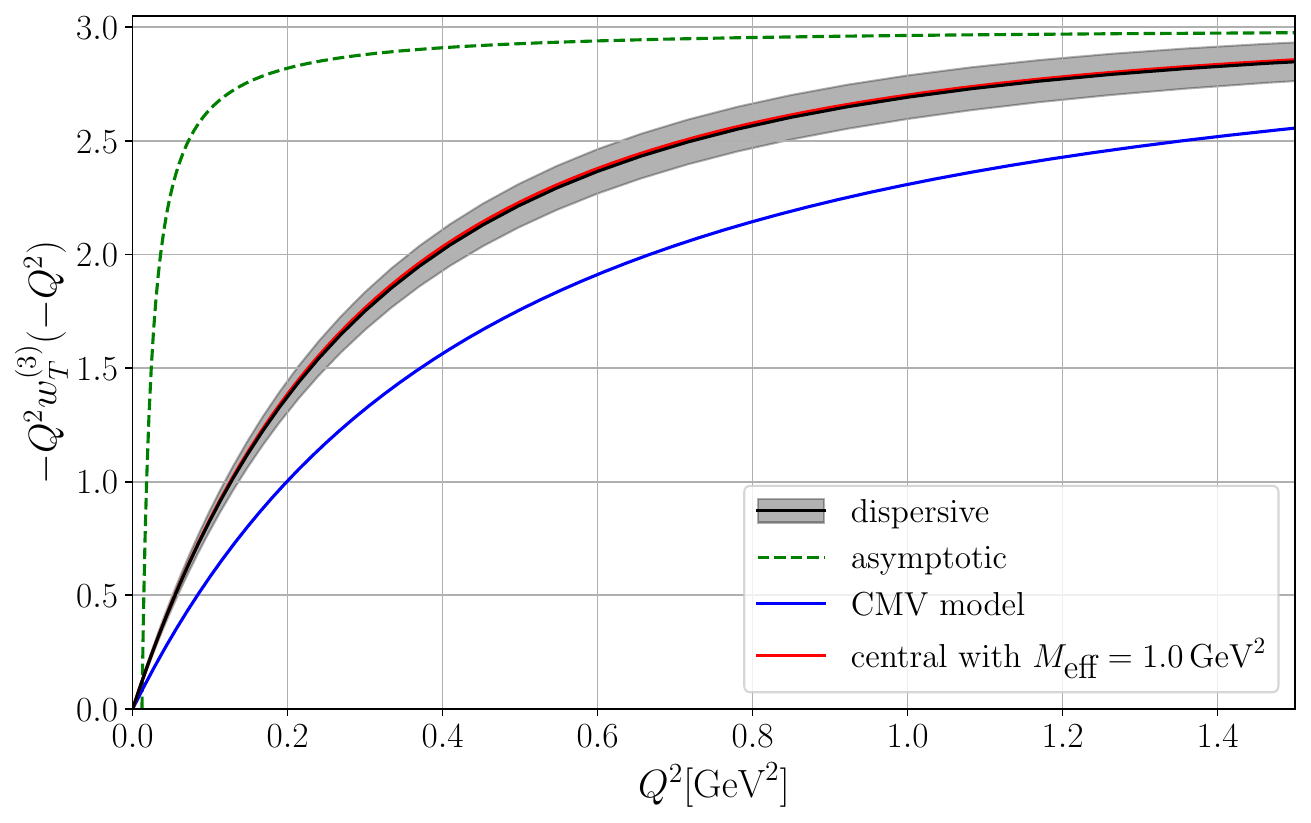}
	\caption{Comparison of the dispersive analysis of
	Ref.~\cite{Ludtke:2024ase} for $w_{L,T}^{(3)}$ to the Czarnecki--Marciano--Vainshtein (CMV) model~\cite{Czarnecki:2002nt,Melnikov:2003xd}, the asymptotic form~\eqref{wLT_anomaly}, and our description~\eqref{wLT_3}.}
	\label{fig:wL_wT}
\end{figure}

While the leading constraints for $w_L^{(3)}$ already agree with Eq.~\eqref{wL_mixed}, the remaining ones
are most easily implemented by supplementing this representation by a pole at $\Mr^2$, as dominating the momentum dependence of the pion TFF, and an effective pole at $\Meff^2$, to describe the transition to the asymptotic region. This procedure leads to
\begin{align}
\label{wLT_3}
 w_L^{(3)}(q^2)&=\frac{6(q^2)^2+6q^2\big(\kappa_\text{OPE}^{(3)}-\mpi^2-\Mr^2-\Meff^2\big)-\mpi^2\Mr^2\Meff^2w_L^{(3)}(0)}{(q^2-\mpi^2)(q^2-\Mr^2)(q^2-\Meff^2)},\notag\\
 w_T^{(3)}(q^2)&=\frac{3(q^2)^2+3q^2\big(\kappa_\text{OPE}^{(3)}-\Ma^2-\Mr^2-\Meff^2\big)-\Ma^2\Mr^2\Meff^2w_T^{(3)}(0)}{(q^2-\Ma^2)(q^2-\Mr^2)(q^2-\Meff^2)}.
\end{align}
Comparing $w_{L,T}^{(3)}(q^2)$ from Eq.~\eqref{wLT_3} to Ref.~\cite{Ludtke:2024ase}, see Fig.~\ref{fig:wL_wT}, we observe perfect agreement for $\Meff=1.5\GeV$ and $\Meff=1.0\GeV$, respectively, and will use the same values below in Eq.~\eqref{wLT08} as well. Figure~\ref{fig:wL_wT} also illustrates the improvement over the previously used model from Refs.~\cite{Czarnecki:2002nt,Melnikov:2003xd}.

Generalizing Eq.~\eqref{wLT_3} to the remaining flavor components,
we do not attempt to separate octet and singlet, but instead implement the constraints
\begin{align}
 w_L^{(0,8)}(0)&=-4\pi^2\sum_{P=\eta,\eta'}\frac{(1+b_P M_P^2)F_{P\gamma\gamma}\bar F^P_\text{asym}}{M_P^2},\notag\\
 w_T^{(0,8)}(0)&=-4\pi^2\sum_{P=\eta,\eta'}b_P F_{P\gamma\gamma}\bar F^P_\text{asym},\notag\\
 w_L^{(0,8)}(q^2)\Big|_\text{OPE}&=2w_T^{(0,8)}(q^2)\Big|_\text{OPE}=\frac{6}{q^2}\bigg[1+\frac{\kappa_\text{OPE}^{(0,8)}}{q^2}+\Order\Big(\big(q^2\big)^{-2}\Big)\bigg],
\end{align}
where $b_\eta = 1.833(41)\GeV^{-2}$, $b_{\eta'}=1.493(32)\GeV^{-2}$~\cite{Holz:2024lom}, and
\beq
\kappa_\text{OPE}^{(0,8)}=\frac{8\pi^2(25\hat m\hat X+2m_s X_s)}{81e}\simeq 0.043 \GeV^2.
\eeq
Keeping $\eta$, $\eta'$ and $f_1$, $f_1'$ contributions in the relative weights implied by Eqs.~\eqref{wL_mixed} and~\eqref{wT_axial}, we obtain
\begin{align}
\label{wLT08}
  w_L^{(0,8)}(q^2)&=\frac{6(q^2)^2+6q^2\big(\kappa_\text{OPE}^{(0,8)}-\xi_\eta M_\eta^2-\xi_{\eta'}M_{\eta'}^2-\Mr^2-\Meff^2\big)-M_{\eta\eta'}^2\Mr^2\Meff^2w_L^{(0,8)}(0)}{(q^2-\Mr^2)(q^2-\Meff^2)}\notag\\
  &\times\sum_{P=\eta,\eta'}\frac{\xi_P}{q^2-M_P^2}
  ,\notag\\
  w_T^{(0,8)}(q^2)&=\frac{3(q^2)^2+3q^2\big(\kappa_\text{OPE}^{(0,8)}-\xi_{f_1} \Mf^2-\xi_{f_1'}\Mfp^2-\Mr^2-\Meff^2\big)-M_{f_1f_1'}^2\Mr^2\Meff^2w_T^{(0,8)}(0)}{(q^2-\Mr^2)(q^2-\Meff^2)}\notag\\
  &\times\sum_{A=f_1,f_1'}\frac{\xi_A}{q^2-M_A^2}
  ,\notag\\
  \xi_P&=\frac{F_{P\gamma\gamma}\bar F^P_\text{asym}}{\sum_{P'=\eta,\eta'}F_{P'\gamma\gamma}\bar F^{P'}_\text{asym}},\qquad
  M_{\eta\eta'}^2=\frac{M_\eta^2 M_{\eta'}^2}{\xi_\eta M_{\eta'}^2+\xi_{\eta'}M_{\eta}^2},\notag\\
  \xi_A&=\frac{\frac{F_{A}^\text{eff}}{M_{A}}\F_2^{A}(0,0)}{\sum_{A'=f_1,f_1'}\frac{F_{A'}^\text{eff}}{M_{A'}}\F_2^{A'}(0,0)},\qquad
  M_{f_1f_1'}^2=\frac{\Mf^2 \Mfp^2}{\xi_{f_1} \Mfp^2+\xi_{f_1'}\Mf^2}.
\end{align}
Given that all the corresponding TFFs depend on $M_\phi$ as well, one could try to impose further flavor constraints by exploiting the mass gap between $\Mr$ and $M_\phi$, but given that the dominant uncertainty derives from the transition to the asymptotic region anyway, a meaningful improvement of Eq.~\eqref{wLT08} would require a dedicated dispersive calculation.

\section{Effective poles}
\label{app:effective_poles}

To assess potential tails of higher intermediate states at small virtualities, we study the impact of effective-pole contributions to the HLbL tensor that could subsume such corrections. In the main text, we show the impact of adding such poles to the matching between the sum of exclusive states and the pQCD description, here, we provide the details of their construction.

In general, to improve the matching to SDCs, it is most efficient to consider a single effective pole with TFFs interpreted in triangle kinematics for HLbL scattering~\cite{Ludtke:2023hvz}, that is, a pseudoscalar pole for $\hat \Pi_{1\text{--}3}$ and an axial-vector one for the other scalar functions, e.g.,
\begin{align}
\hat \Pi_1^\text{eff}&=\frac{F_{P\gamma^*\gamma^*}(q_1^2,q_2^2)F_{P\gamma^*\gamma^*}(M_P^2,0)}{q_3^2-M_P^2},\notag\\
\hat \Pi_4^\text{eff}
&=\frac{(q_1^2+q_3^2-\mA^2)\F_2^A(\mA^2,0)\big[2\F_1^A(q_1^2,q_3^2)+\F_3^A(q_1^2,q_3^2)\big]}{2\mA^4(q_2^2-\mA^2)}+\big(q_1^2\leftrightarrow q_2^2\big),
\end{align}
where an easy way to implement the asymptotic behavior proceeds via a quark-model-inspired form~\cite{Schuler:1997yw,Hoferichter:2020lap}, i.e.,
\begin{align}
\frac{F_{P\gamma^*\gamma^*}(q_1^2,q_2^2)}{F_{P\gamma^*\gamma^*}(0,0)}&=\frac{\Lambda_P^2}{\Lambda_P^2-q_1^2-q_2^2},\notag\\
\frac{\F_2^A(q_1^2,q_2^2)}{\F_2^A(0,0)}&=\frac{\F_3^A(q_1^2,q_2^2)}{\F_3^A(0,0)}=\bigg(\frac{\Lambda_A^2}{\Lambda_A^2-q_1^2-q_2^2}\bigg)^2,\qquad
\F_1^A(q_1^2,q_2^2)=0,
\end{align}
with normalizations and scales $\Lambda_P$, $\Lambda_A$ to be determined. Matching the scalar functions in the symmetric asymptotic limit $q_1^2=q_2^2=q_3^2$ to pQCD~\cite{Colangelo:2019uex,Bijnens:2021jqo}, we obtain
\begin{align}
\label{matching_sym}
 F_{P\gamma^*\gamma^*}(0,0)F_{P\gamma^*\gamma^*}(M_P^2,0)\Lambda_P^2&=\frac{8}{9\pi^2}\bigg(1-c_1\frac{\alpha_s}{\pi}\bigg),\\
 \frac{\F_2^A(0,0)\F_2^A(\mA^2,0)\Lambda_A^4}{\mA^4}&=\begin{cases}
           \frac{16}{243\pi^2}\Big[33-16\sqrt{3}\,\text{Cl}_2\big(\frac{\pi}{3}\big)\Big]\Big(1-c_{i}\frac{\alpha_s}{\pi}\Big), &\quad i=4,7,\\
           -\frac{64}{81\pi^2}\Big[3-2\sqrt{3}\,\text{Cl}_2\big(\frac{\pi}{3}\big)\Big]\Big(1-c_{17}\frac{\alpha_s}{\pi}\Big), &\quad i=17,\\
           \frac{32}{729\pi^2}\Big[15-4\sqrt{3}\,\text{Cl}_2\big(\frac{\pi}{3}\big)\Big]\Big(1-c_{39}\frac{\alpha_s}{\pi}\Big), &\quad i=39,
	\end{cases}\notag
\end{align}
with Clausen function $\text{Cl}_2(x)=\Im\text{Li}_2(e^{ix})$ and coefficients
\beq
c_{1,4,7,17,39}\simeq \big\{0.86,0.30,0.68,2.15,0.85\big\}.
\eeq
Remarkably, the four axial-vector conditions almost coincide, the numerical coefficients on the right-hand side being $\{31.4,30.0,31.4,32.1\}\times 10^{-3}$ when including $\alpha_s$ corrections (at scale $\mu=1.5\GeV$). $\hat \Pi_{54}$ vanishes identically in the symmetric limit.

Matching instead in the limit $|q_3^2| \ll |q_1^2| = |q_2^2|$, the analog conditions become
\begin{align}
\label{matching_asym}
 F_{P\gamma^*\gamma^*}(0,0)F_{P\gamma^*\gamma^*}(M_P^2,0)\Lambda_P^2&=\frac{4}{3\pi^2}\bigg(1-\frac{\alpha_s}{\pi}\bigg),\notag\\
 \frac{\F_2^A(0,0)\F_2^A(\mA^2,0)\Lambda_A^4}{\mA^4}&=\frac{8}{9\pi^2}+\Order\Big(\frac{\alpha_s}{\pi}\Big),
\end{align}
where the $\alpha_s$ corrections to $\hat \Pi_{4,7,17,39}$ take a more complicated form due to the appearance of logarithms $\log\frac{q_3^2}{\hat q^2}$.
Further, we discard the condition from $\hat\Pi_{54}$ because of the mismatch by a factor two already observed in App.~\ref{app:VVA_matching}.
Comparing the coefficients in Eqs.~\eqref{matching_sym} and~\eqref{matching_asym}, we see that the asymmetric matching requires a somewhat larger coefficient, by factors $1.5$ and $\simeq 2.8$ in the pseudoscalar and axial-vector cases, respectively. The latter may be overestimated, however, due to $\alpha_s$ corrections in Eq.~\eqref{matching_asym} (comparing only the leading terms, the mismatch reduces to $\simeq 2.5$). To simplify these matching conditions, we define dimensionless effective couplings
\beq
g_P^\text{eff}\equiv M_P^2
F_{P\gamma^*\gamma^*}(0,0)F_{P\gamma^*\gamma^*}(M_P^2,0),\qquad
g_A^\text{eff}\equiv\F_2^A(0,0)\F_2^A(\mA^2,0),
\eeq
in terms of which
\begin{align}
\label{geff}
 &\text{symmetric:} & g_P^\text{eff}&\simeq 0.081\Big(\frac{M_P}{\Lambda_P}\Big)^2, & g_A^\text{eff}&\simeq 0.031\Big(\frac{M_A}{\Lambda_A}\Big)^4,\notag\\
 &\text{asymmetric:} & g_P^\text{eff}&\simeq 0.121\Big(\frac{M_P}{\Lambda_P}\Big)^2, & g_A^\text{eff}&\simeq 0.090\Big(\frac{M_A}{\Lambda_A}\Big)^4.
\end{align}

\bibliographystyle{apsrev4-1_mod_2}
\bibliography{amu}
	
\end{document}